\documentstyle{article}
\makeatletter
\def\emline#1#2#3#4#5#6{%
       \put(#1,#2){\special{em:moveto}}%
       \put(#4,#5){\special{em:lineto}}}
\def\newpic#1{}

\def\hybrid{\topmargin 0pt      \oddsidemargin 0pt
        \headheight 0pt \headsep 0pt
       \voffset-1cm
       \textwidth 6.5in        
       \textheight 9in         
        \marginparwidth 0.0in
        \parskip 5pt plus 1pt   \jot = 1.5ex}
\catcode`\@=11
\def\marginnote#1{}

\newcount\hour
\newcount\minute
\newtoks\amorpm
\hour=\time\divide\hour by60
\minute=\time{\multiply\hour by60 \global\advance\minute by-\hour}
\edef\standardtime{{\ifnum\hour<12 \global\amorpm={am}%
        \else\global\amorpm={pm}\advance\hour by-12 \fi
        \ifnum\hour=0 \hour=12 \fi
        \number\hour:\ifnum\minute<10 0\fi\number\minute\the\amorpm}}
\edef\militarytime{\number\hour:\ifnum\minute<10 0\fi\number\minute}

\def\draftlabel#1{{\@bsphack\if@filesw {\let\thepage\relax
   \xdef\@gtempa{\write\@auxout{\string
      \newlabel{#1}{{\@currentlabel}{\thepage}}}}}\@gtempa
   \if@nobreak \ifvmode\nobreak\fi\fi\fi\@esphack}
        \gdef\@eqnlabel{#1}}
\def\@eqnlabel{}
\def\@vacuum{}
\def\draftmarginnote#1{\marginpar{\raggedright\scriptsize\tt#1}}

\def\draftlabel#1{{\@bsphack\if@filesw {\let\thepage\relax
   \xdef\@gtempa{\write\@auxout{\string
      \newlabel{#1}{{\@currentlabel}{\thepage}}}}}\@gtempa
   \if@nobreak \ifvmode\nobreak\fi\fi\fi\@esphack}
        \gdef\@eqnlabel{#1}}
\def\@eqnlabel{}
\def\@vacuum{}
\def\draftmarginnote#1{\marginpar{\raggedright\scriptsize\tt#1}}

\def\draft{\oddsidemargin -.5truein
        \def\@oddfoot{\sl preliminary draft \hfil
        \rm\thepage\hfil\sl\today\quad\militarytime}
        \let\@evenfoot\@oddfoot \overfullrule 3pt
        \let\label=\draftlabel
        \let\marginnote=\draftmarginnote
   \def\@eqnnum{(\theequation)\rlap{\kern\marginparsep\tt\@eqnlabel}%
\global\let\@eqnlabel\@vacuum}  }

\def\numberbysection{\@addtoreset{equation}{section}
        \def\theequation{\thesection.\arabic{equation}}}

\def\underline#1{\relax\ifmmode\@@underline#1\else
        $\@@underline{\hbox{#1}}$\relax\fi}

\def\titlepage{\@restonecolfalse\if@twocolumn\@restonecoltrue\onecolumn
     \else \newpage \fi \thispagestyle{empty}\c@page\z@
        \def\thefootnote{\fnsymbol{footnote}} }

\def\endtitlepage{\if@restonecol\twocolumn \else  \fi
        \def\thefootnote{\arabic{footnote}}
        \setcounter{footnote}{0}}  
\relax

\numberbysection
\hybrid

\def\beq{\begin{equation}}
\def\eeq{\end{equation}}

\def\t{\tau}
\def\l{\lambda}
\def\a{\alpha}
\def\b{\beta}

\begin{document}

\begin{titlepage}
\title{
\rightline{\normalsize ITEP-TH-45/97}
Hidden quantum $R$-matrix
in discrete time classical Heisenberg magnet}

\author{A.Zabrodin\thanks{Joint Institute of
Chemical Physics, Kosygina str. 4, 117334,
Moscow, Russia and ITEP, 117259, Moscow, Russia;
e-mail: zabrodin@heron.itep.ru}}
\date{October 1997}
\maketitle

\begin{abstract}

We construct local $M$-operators
for an integrable discrete time version of the
classical Heisenberg magnet by convolution of the twisted
quantum trigonometric 4$\times$4
$R$-matrix with certain vectors in its "quantum" space.
Components of the vectors are identified
with $\tau$-functions of the model.
Hirota's bilinear formalism is extensively used.
The construction
generalizes the known representation of $M$-operators
in continuous time models in terms of Lax operators
and classical $r$-matrix.

\end{abstract}


\end{titlepage}

\section{Introduction}

The unified treatment \cite{FadTakh} of
non-linear soliton equations as
hamiltonian systems having enough number of conserved
quantities in involution is based on the (classical) $r$-matrix.
Its role is to provide a universal
form of Poisson brackets for elements of Lax operators.
An alternative though less popular point of view
on the $r$-matrix (which we are going to follow)
comes from the Zakharov-Shabat
approach \cite{FadTakh},\,\cite{ZMNP}
which consists in representing
soliton equations as 2D zero curvature conditions
for a pair of matrix functions (called $L$ and $M$ operators)
depending on a spectral parameter. The $r$-matrix works there
as a machine producing $M$-operators from $L$-operators.

In the paper \cite{Zsg},
for the simplest example of the lattice sine-Gordon (SG) model,
we have found a similar machine in the completely discrete set-up,
i.e., in lattice integrable models with discrete time.
Remarkably enough, it appears to be a {\it quantum $R$-matrix}
with the "quantum" parameter $q$ related to the time lattice
spacing. The formula for local
$M$-operators in the discrete case has as simple structure
as in the continuous one (see (\ref{r6}) below)
with the $R$-matrix in place of the $r$-matrix.
This comes as the result of a computation.
At the moment we do not try to
"explain" why a typical quantum $R$-matrix takes part in
a purely classical problem.

In this paper we show that the construction still works for a
much more general model --  partially anisotropic
lattice Heisenberg magnet (HM) in
discrete time. (This model contains discrete versions
of SG, KdV and other equations as special cases.)
Although the final result looks
like a straightforward generalization
of the one for the lattice SG model,
the derivation gets much more
involved and some important modifications are necessary.

Let us recall the $r$-matrix construction of $M$-operators
for continuous flows.
Let ${\cal L}_l (z)$ be a classical
2$\times$2 $L$-operator on 1D lattice with the periodic
boundary condition ${\cal L}_{l+N}(z)={\cal L}_{l}(z)$;
$z$ is the spectral parameter. For the moment we assume the
ultralocality -- Poisson brackets between elements of
the ${\cal L}_l (z)$ at different sites are zero.
The monodromy matrix ${\cal T}_{l}(z)$ is ordered product
of $L$-operators along the lattice from the site $l$ to
$l+N-1$:
\beq
{\cal T}_{l}(z) =
{\cal L}_{l+N-1}(z)\ldots \,
{\cal L}_{l+1}(z) {\cal L}_{l}(z)\,.
\label{r1}
\eeq
Hamiltonians of commuting flows are obtained by
expanding $\mbox{log}T(z)$ in $z$, where
$T(z)=\mbox{tr}\, {\cal T}_{l}(z)$
does not depend on $l$ due to the periodic boundary condition.
All these flows admit a zero curvature representation
with the generating function of
$M$-operators given by \cite{Sklyanin},\,\cite{FadTakh}
\beq
M_{l}(z;w) =T^{-1}(w)\, \mbox{tr}_{1}
\left [ r\big (z/w \big )
({\cal T}_{l}(w)\otimes I) \right ].
\label{r3}
\eeq
Here $r(z)$ is a (classical) 4$\times$4 $r$-matrix,
acting in the tensor product of two 2-dimensional spaces,
$\mbox{tr}_{1}$ means trace in the first space,
$I$ is the unity matrix.
Expanding the r.h.s. of eq.\,(\ref{r3})
in $w$, one gets $M$-operators depending on the spectral
parameter $z$.
From the hamiltonian point of view,
the zero curvature condition follows from Poisson
brackets for elements of the $L$-operator.
A similar $r$-matrix formula exists for
$M$-operators in non-ultralocal models, though, in this
case the $r$-matrix is not skew-symmetric.
In general, $M_{l}(z;w)$ is non-local.

It is well known \cite{IK},\,\cite{FadTakh}
how to construct local $M$-operators from the generating
function. Suppose there exists $z_0$
such that $ \det {\cal L}_l (z_0 )=0$ for any $l$. This means that
${\cal L}_{l}(z_0 )$ is
a 1-dimensional projector:
\beq
{\cal L}_l (z_0 )=\frac{\big | \a (l)\big >
\big < \b (l) \big |}{P(l)}\,,
\label{r4}
\eeq
where
\beq
\big | \a \big >=
\left (\! \begin{array}{c} \a _{1}\\ \a _{2}\end{array}
\! \right ),
\;\;\;\;\;\;\;\;
\big < \b \big |=\big ( \b _{1},\, \b _{2}\big )
\label{r5}
\eeq
are two-component vectors and $P(l)$ is a scalar
normalization factor.
Components of the vectors as well as the $P(l)$
depend on dynamical variables.
It is easy to see that $M_{l}(z;z_0 )$ is a local quantity:
\beq
M_l (z)\equiv  M_{l}(z; z_0 )=
\frac{ \big < \b (l)
\big | r(z/z_0 )\big |\a (l-1) \big >}
{\big < \b (l)\big | \a (l-1)\big >}
\label{r6}
\eeq
(note that the normalization factor cancels).
The scalar product is taken in the first space only, so the result
is a 2$\times$2 matrix with the spectral parameter $z$.
It obeys the zero curvature condition
\beq
\partial _{t}{\cal L}_{l}(z)=
M_{l+1}(z){\cal L}_{l}(z)-
{\cal L}_{l}(z)M_{l}(z)\,.
\label{czc}
\eeq

Let us outline the results of this work.
We consider the completely
discretized partially anisotropic ($XXZ$) HM
and represent $M$ and $L$ operators
${\cal M}_l (z)$, ${\cal L}_l (z)$
by formulas of the type (\ref{r6}):

\beq
{\cal M}_l (z)=
\frac{ \big < \b (l)
\big | R(z/z_0 ;q, \xi )\big |\check \b (l-1) \big >}
{\big < \b (l)\big | \a (l-1)\big >}\,,
\label{r8}
\eeq

\beq
{\cal L}_l (z)=
\frac{ \big < \b (l)
\big | R(z/z_0 ;q, \xi ' )\big |\a (l) \big >}
{\big < \b (l)\big | \a (l-1)\big >}\,.
\label{r9}
\eeq

\noindent
In the r.h.s., $R(z;q, \xi )$, $R(z;q, \xi ')$
are {\it quantum} 4$\times$4 $R$-matrices (to be
specified below) with the "quantum" parameter $q$
and the Drinfeld twist parameters $\xi , \xi '$ related
to the space-time lattice.
The vectors
$\big | \a (l) \big >$, $\big | \b (l) \big >$ are
{\it the same} as in eq.\,(\ref{r6}),
$$\big |\check \b (l) \big >\equiv
\left ( \begin{array}{cc}0&(-\xi \xi ')^{\frac{1}{2}}
\\ (-\xi \xi ')^{-\frac{1}{2}}&0 \end{array} \right )
\big | \b (l) \big >.$$
It is worth noting here that the
continuous and discrete time models
have a common $L$-operator; formula (\ref{r9}) gives
its $R$-matrix representation.

To visualize these formulas, it is convenient to use
the language of the
algebraic Bethe ansatz \cite{FT},\,\cite{book}.
The scalar product is taken in the "quantum" (vertical) space,
so one gets a 2$\times$2 matrix in
the "auxiliary" (horizontal) space:

\begin{center}
\special{em:linewidth 0.4pt}
\unitlength 0.6mm
\linethickness{0.4pt}
\begin{picture}(90.00,59.00)(25.00,10.00)
\emline{60.00}{40.00}{1}{90.00}{40.00}{2}
\emline{75.00}{55.00}{3}{75.00}{25.00}{4}
\put(75.00,59.00){\makebox(0,0)[cc]{$\big <\beta \big |$}}
\put(75.00,20.00){\makebox(0,0)[cc]{$\big |\alpha \big >$}}
\put(52.00,40.00){\makebox(0,0)[rc]
{$\big < \beta \big | R(z)\big |\alpha \big >\, =\, $}}
\end{picture}
\end{center}

\vspace{-0.4cm}

The $M$-operator (\ref{r8}) generates shifts of a discrete
time variable $m$.
The discrete zero curvature condition
\beq
{\cal M}_{l+1,m}(z){\cal L}_{l,m}(z)
={\cal L}_{l,m+1}(z){\cal M}_{l,m}(z)
\label{dzc}
\eeq
gives rise to
the HM equations of motion in discrete space-time.

The key ingredient of the construction is to
pass from the initial dynamical variables to the pair of vectors
$\big | \a (l) \big >$, $\big | \b (l) \big >$.
In the papers \cite{IK}
on exact lattice regularization of integrable models
components of these vectors were expresed in terms of
canonical variables of the model. Those formulas looked
quite complicated and were hardly considered as something
instructive. Here we shed some light on their meaning.
Using equations of motion of the completely
discretized model (derived from the discrete zero curvature
condition (\ref{dzc})), we show that (suitably normalized)
components of the vectors
$\big | \a (l) \big >$,
$\big | \b (l) \big >$
are {\it $\t$-functions}.

The $\t$-function is
one of the most fundamental objects of the theory
(see e.g. \cite{JM},\,\cite{SW}). It is worth stressing that
quantum $R$-matrices in the classical discrete
problems can hardly be noticed until one reformulates
the non-linear equations and
elements of the $L$-$M$ pair entirely in terms
the $\t$-functions. That is why we are to make a long
excursion into Hirota's bilinear
formalism \cite{HirotaKdV}-\cite{Hirota} (Sect.\,2).
Our guiding principle is Miwa's interpretation \cite{Miwa} of
the discrete time flows, in which discretized
integrable equations are treated as members
of the same infinite hierarchy
as their continuous counterparts.
A general method to produce discrete soliton equations
was developed in \cite{DJM},
where, in particular, the discrete
isotropic ($XXX$) HM model was proposed.
Following these ideas, we deal with
bilinear form of the discrete $XXZ$ HM model.
The matrix $L$-$M$-pair for
the latter is derived in Sect.\,3
from basic linear problems for a scalar wave
function \cite{Hirota},\,\cite{DJM},\,\cite{SS}.

Like in the SG model,
there are two lattice versions of the classical HM: the
discrete HM model on a space-time lattice \cite{DJM}
and the model on a space lattice with continuous time
introduced in \cite{Sklfun}
for the more general $XYZ$-case.
It is interesting to note that both of them were
proposed back in 1982 but
the ideas underlying one and another
seemed to be "orthogonal" and did not
intersect until the very recent time\footnote{On the
quantum level, these ideas were partially linked together
in \cite{FadVol}-\cite{BBR}, where
{\it quantum} SG model in discrete space-time
(a quantization of Hirota's discrete SG equation)
was constructed; a generalization to affine Toda
field theories on the lattice was suggested in \cite{KR}.}.
A motivation of this work was just an attempt to
understand one of the models in terms of the other one.

The manner of exposition here is different from the one
in \cite{Zsg}, where we started with
either Faddeev-Volkov or Izergin-Korepin
$L$-operator for the lattice SG model
in terms of lattice fields and
passed to the $\t$-function by means of a special
substitution.
In this paper we start directly from the
$\t$-function (and bilinear equations it obeys)
rather than from lattice spin variables. The reason
is twofold. First, the equations of motion in terms of
lattice spin variables are too complicated and, anyway,
are of no practical use for us.
Second, in the discrete case there is
no canonical way to introduce spin variables whereas
Hirota's bilinear formalism is gauge-invariant
and free of artifacts. Having this in mind, let us remark
that by {\it Heisenberg model} we mean here nothing more
than a convenient name for the properly reduced
2D Toda lattice hierarchy \cite{UT}. (Actually we deal
with a model which is slightly more general than
the $XXZ$ HM itself.) In this sense a name
like {\it partially anisotropic chiral field model}
would be also appropriate. Needless to say that in the
discrete set-up the specific features of each of the two
models become irrelevant \cite{JM}. What {\it is} relevant
is the type of reduction of the 2D Toda hierarchy that is
the same in both cases.

Treated as a {\it hierarchy},
the bilinear equations imply infinitely many discrete variables
(commuting flows) from the very beginning.
Generally speaking, any two of them could be chosen as
space-time coordinates. However, for our purpose
we need a distinguished space-time lattice.
In Sect.\,4 we introduce such a lattice
with coordinates denoted by
$l,m$ and derive the matrix $L$-$M$-pair
for translations $l\to l+1$, $m\to m+1$.
These $L$ and $M$ operators
are represented in the form (\ref{r9}), (\ref{r8}) in
Sect.\,5. In Sect.\,6 we discuss the continuous time limit
of the construction and show how eq.\,(\ref{r6})
is reproduced. As it was already mentioned, the $r$-matrix
in that formula turns out to be not necessarily skew-symmetric.
This fact is a signal of non-ultralocality of the model
in general case. Sect.\,7 is a conclusion where we point out
a few unsolved problems motivated by our results.

\section{Hirota's bilinear formalism}

In this section we present
the results of the papers \cite{Miwa},\,\cite{DJM}
in the form convenient for our purpose.
We illustrate formulas by the
graphical representation of discrete flows
suggested in the review \cite{Zab} to which we refer
for more details.

\subsection{General form of 3-term bilinear equations}

The key object of Hirota's approach is {\it $\t$-function}.
The $\t$-function $\t = \t (a,b,c, \ldots)$ (as a function
of the discrete variables $a,b,c, \ldots $) obeys a number
of bilinear partial difference equations.
To each discrete variable $a$
a complex parameter $\l _{a}\in {\bf C}$
("Miwa's variable") is associated.
One may think of the $\l _a$ as a continuous "number"
which marks the discrete flow.
For any two discrete flows $a,b$ we put
$$
\l _{ab}\equiv \l _a -\l _b \,,
\;\;\;\;\;\;
\l _{ab}=-\l _{ba}\,.
$$

Let us recall how to compose bilinear equations.
Each triplet $\{ a,b,c\}$ of discrete flows
gives rise to a 3-term bilinear
equation \cite{Hirota},\,\cite{Miwa}
for the $\t$:

\begin{eqnarray}
&&\l _{bc}\t (a+1, b,c)\t (a, b+1, c+1)
\nonumber \\
&+&\l _{ca}\t (a, b+1,c)\t (a+1, b, c+1)
\nonumber \\
&+&\l _{ab}\t (a, b,c+1)\t (a+1, b+1, c)=0\,.
\label{tripl}
\end{eqnarray}

\noindent
All other variables which the $\t$ may depend on enter
this equation as parameters.
Each quadruplet $\{ a,b,c,d \}$
gives rise to another 3-term bilinear
equation:

\begin{eqnarray}
&&\l _{ad}\l _{bc}\t (a+1, b,c, d+1)\t (a, b+1, c+1,d)
\nonumber \\
&+&\l _{bd}\l _{ca}\t (a, b+1,c, d+1)\t (a+1, b, c+1,d)
\nonumber \\
&+&\l _{cd}\l _{ab}\t (a, b,c+1, d+1)\t (a+1, b+1, c,d)=0\,.
\label{quadrupl}
\end{eqnarray}

\noindent
{\bf Remark}\,\,Links between
eqs.\,(\ref{tripl}), (\ref{quadrupl})
exist in both directions.
On the one hand, eq.\,(\ref{tripl}) is a particular case
of (\ref{quadrupl}) when $\l _{d} \to \infty$. (According
to \cite{Miwa}, this limit means that the dependence
on $d$ in the $\t$ disappears.) On the other hand,
eq.\,(\ref{quadrupl}), though linearly independent
of eqs.\,(\ref{tripl}), is an {\it algebraic consequense} of
equations of the type (\ref{tripl}) written for
the triplets $\{a,b,c\}$, $\{a,b,d\}$ and $\{a,c,d\}$.
In this sense all what we are going to derive in the
sequel follows already from eqs.\,(\ref{tripl}).

\subsection{Basic bilinear equations for the discrete HM model}

Let us consider five discrete flows and denote
the corresponding variables by $p$, $p'$, $n$, $u$, $v$.
We denote the $\t$-function by
$\t (p,p' , n,u,v) \equiv \t ^{p,p'}_{n}(u,v)$.
In the latter notation the $u,v$ are separated from the
others because they
will play the role of chiral space-time coordinates.
We call them {\it chiral variables}.

Bilinear equations for the $\t$-function of the $XXZ$ HM model
are obtained from (\ref{tripl}), (\ref{quadrupl})
by imposing the reduction
\beq
\t ^{p+1,p'+1}_{n}(u,v)
=\t ^{p,p'}_{n+1}(u,v)\,,
\label{b1}
\eeq
so we are left with four independent variables.
Let us choose them to be $p$, $n$, $u$, $v$.

\noindent
{\bf Remark}\,\,The discrete KdV equation is a particular case
$\l _n =\infty$, so the dependence on $n$ disappears.
The discrete SG model corresponds to further specification
$\l _{p'} =\l _{p}$.

Using the graphical representation introduced in \cite{Zab},
we can display the above configuration of discrete flows
in the figure (the graph of flows):

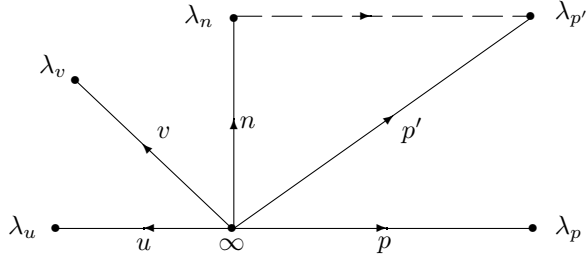
\begin{center}
\special{em:linewidth 0.4pt}
\unitlength 0.8mm
\linethickness{0.4pt}
\begin{picture}(93.67,56.00)
\emline{10.33}{20.00}{1}{90.00}{20.00}{2}
\put(90.00,20.00){\circle*{1.33}}
\put(10.67,20.00){\circle*{1.33}}
\put(40.00,20.00){\circle*{1.33}}
\emline{40.33}{20.00}{3}{40.33}{55.00}{4}
\put(40.33,55.00){\circle*{1.33}}
\emline{40.33}{20.00}{5}{89.67}{55.00}{6}
\put(89.67,55.33){\circle*{1.33}}
\emline{40.00}{20.00}{7}{14.00}{44.67}{8}
\put(14.00,44.67){\circle*{1.33}}
\emline{40.00}{55.33}{9}{45.00}{55.33}{10}
\emline{46.67}{55.33}{11}{51.67}{55.33}{12}
\emline{53.33}{55.33}{13}{58.33}{55.33}{14}
\emline{60.00}{55.33}{15}{65.00}{55.33}{16}
\emline{66.67}{55.33}{17}{71.67}{55.33}{18}
\emline{73.33}{55.33}{19}{78.33}{55.33}{20}
\emline{80.00}{55.33}{21}{85.00}{55.33}{22}
\emline{86.67}{55.33}{23}{89.33}{55.33}{24}
\put(66.75,38.75){\vector(4,3){0.2}}
\emline{66.14}{38.30}{25}{66.75}{38.75}{26}
\put(25.29,33.96){\vector(-1,1){0.2}}
\emline{26.14}{33.11}{27}{25.29}{33.96}{28}
\put(40.33,38.14){\vector(0,1){0.2}}
\emline{40.33}{37.31}{29}{40.33}{38.14}{30}
\put(65.65,20.01){\vector(1,0){0.2}}
\emline{64.44}{20.01}{31}{65.65}{20.01}{32}
\put(25.48,20.01){\vector(-1,0){0.2}}
\emline{26.13}{20.01}{33}{25.48}{20.01}{34}
\put(62.96,55.34){\vector(1,0){0.2}}
\emline{62.05}{55.34}{35}{62.96}{55.34}{36}
\put(7.67,20.33){\makebox(0,0)[rc]{$\lambda _{u}$}}
\put(10.33,47.00){\makebox(0,0)[cc]{$\lambda _{v}$}}
\put(37.00,55.33){\makebox(0,0)[rc]{$\lambda _{n}$}}
\put(93.67,55.67){\makebox(0,0)[lc]{$\lambda _{p'}$}}
\put(93.67,20.00){\makebox(0,0)[lc]{$\lambda _{p}$}}
\put(40.00,17.33){\makebox(0,0)[cc]{$\infty$}}
\put(25.33,17.33){\makebox(0,0)[cc]{$u$}}
\put(65.33,17.33){\makebox(0,0)[cc]{$p$}}
\put(42.67,37.67){\makebox(0,0)[cc]{$n$}}
\put(28.67,36.00){\makebox(0,0)[cc]{$v$}}
\put(70.00,36.33){\makebox(0,0)[cc]{$p'$}}
\end{picture}
\end{center}

\vspace{-1.0cm}

\noindent
The dashed line is drown here to indicate that
the reduction (\ref{b1}) is the same as
in the case of the 1D Toda chain in discrete time (a
detailed discussion of this point see in \cite{Zab}).
Indeed, the vector field $\partial _{\bar p}\equiv
-\partial _{n}+\partial _{p}$
defines the flow $\bar p$ corresponding
to the dashed line; in these terms the reduction
acquires the more familiar Toda-like form
$\t ^{p+1, \bar p +1} =\t ^{p, \bar p }$.
This turns into a differential condition
as $\l _{p}\to \infty$, $\l _{p'}\to \l _{n}$.
In this case one gets the isotropic ($XXX$)
HM model in discrete time.

\noindent
{\bf Remark}\,\,Looking at the graph makes it easier
to deal with different discrete flows when there are
many of them.
We refer to \cite{Zab} for the exact meaning of the
graphical elements. Here we only note that keeping in mind
solutions of the finite-gap type (see e.g.
\cite{ZMNP},\,\cite{Kri}),
the reader may think of this figure as drown on a
patch of a Riemann surface with local coordinate
$\l ^{-1}$. Miwa's variables are coordinates of
punctures on the surface.
The lines are then cuts between the punctures,
which give rise to discrete commuting flows
on the Jacobian via the Abel map\footnote{For the general theory
of finite-gap solutions to Hirota's difference equation
see \cite{KWZ}.}.

The condition (\ref{b1}) allows one to get rid of
$p'$ if necessary.
Let us give a list of bilinear equations
which are obtained in this way from
eqs.\,(\ref{tripl}), (\ref{quadrupl}). In front of each equation
we indicate the triplet or quadruplet which it comes from.

The simplest equation involves the variables $p,n$ only:
\beq
\{pp'n\}: \;\;\;\;
\l _{p'n}\t _{n+1}^{p-1}(u,v)\t _{n-1}^{p+1}(u,v)
-\l _{pn}\t _{n}^{p+1}(u,v)\t _{n}^{p-1}(u,v)
=\l _{p'p}\big (\t _{n}^{p}(u,v)\big )^2\,.
\label{pp'n}
\eeq
It is the discrete time 1D Toda chain equation in bilinear
form \cite{Hirota}.
In our case it plays the role of a constraint on the
dynamical variables since the space-time coordinates $u,v$
enter as parameters.

Equations of the next group contain shifts of
$p, n, u$ only:

\begin{eqnarray}
\{pp'u\}: \;\;\;\;\;&&
\l _{p'u}\t _{n}^{p+1}(u,v)\t _{n+1}^{p-1}(u+1,v)\!
-\!\l _{pu}\t _{n+1}^{p-1}(u,v)\t _{n}^{p+1}(u+1,v)
\nonumber\\
&=&\l _{p'p}\t _{n}^{p}(u+1,v)\t _{n+1}^{p}(u,v)\,,
\label{pp'u}
\end{eqnarray}

\begin{eqnarray}
\{pnu\}: \;\;\;\;\;&&
\l _{un}\t _{n}^{p+1}(u,v)\t _{n+1}^{p}(u+1,v)
\!+\!\l _{pu}\t _{n+1}^{p}(u,v)\t _{n}^{p+1}(u+1,v)
\nonumber\\
&=&\l _{pn}\t _{n}^{p}(u+1,v)\t _{n+1}^{p+1}(u,v)\,,
\label{pnu}
\end{eqnarray}

\begin{eqnarray}
\{p'nu\}: \;\;\;\;\;&&
\l _{un}\t _{n}^{p}(u,v)\t _{n}^{p+1}(u+1,v)
\!+\!\l _{p'u}\t _{n}^{p+1}(u,v)\t _{n}^{p}(u+1,v)
\nonumber\\
&=&\!\l _{p'n}\t _{n+1}^{p}(u,v)\t _{n-1}^{p+1}(u+1,v)\,,
\label{p'nu}
\end{eqnarray}

\begin{eqnarray}
\{pp'nu\}: \;\;\;\;\;&&
\l _{pn}\l _{p'u}
\t _{n}^{p+1}(u,v)\t _{n}^{p-1}(u+1,v)
\!-\!\l _{p'n}\l _{pu}
\t _{n+1}^{p-1}(u,v)\t _{n-1}^{p+1}(u+1,v)
\nonumber \\
&=&\l _{p'p}\l _{un}
\t _{n}^{p}(u,v)\t _{n}^{p}(u+1,v)\,.
\label{pp'nu}
\end{eqnarray}

\noindent
Similar equations can be written for $p,n,v$ -- it is enough
to replace $u$ by $v$ everywhere.

\noindent
{\bf Remark}\,\,Combining
eqs.\,(\ref{pp'u}), (\ref{pnu}),
$$
\frac{\mbox{eq.\,(\ref{pp'u})}}
{\l _{p'u}\t _{n+1}^{p-1}(u+1,v)}
-\frac{\mbox{eq.\,(\ref{pnu})}}
{\l _{un}\t _{n+1}^{p}(u+1,v)}\,,
$$
and plugging (\ref{p'nu}) into the l.h.s.,
we obtain eq.\,(\ref{pp'nu}).

Equations involving both coordinates $u,v$ read:

\begin{eqnarray}
\{nuv\}: \;\;\;&&
\l _{un}\t _{n}^{p}(u,v+1)\t _{n+1}^{p}(u+1,v)
\!-\!\l _{vn}\t _{n}^{p}(u+1,v)\t _{n+1}^{p}(u,v+1)
\nonumber \\
&=&\l _{uv}\t _{n}^{p}(u+1,v+1)\t _{n+1}^{p}(u,v)\,,
\label{nuv}
\end{eqnarray}

\begin{eqnarray}
\{pp'uv\}: \;\;\;&&
\l _{p'u}\l _{pv}
\t _{n}^{p+1}(u,v+1)\t _{n+1}^{p-1}(u+1,v)
\!-\!\l _{pu}\l _{p'v}
\t _{n+1}^{p-1}(u,v+1)\t _{n}^{p+1}(u+1,v)
\nonumber \\
&=&\l _{p'p}\l _{uv}
\t _{n}^{p}(u+1,v+1)\t _{n+1}^{p}(u,v)\,,
\label{pp'uv}
\end{eqnarray}

\vspace{0.2cm}

\begin{eqnarray}
\{puv\}: \;\;\; &&
\l _{pv}\t _{n}^{p+1}(u,v+1)\t _{n}^{p}(u+1,v)
\!-\!\l _{pu}\t _{n}^{p+1}(u+1,v)\t _{n}^{p}(u,v+1)
\nonumber \\
&=&\l _{uv}\t _{n}^{p}(u+1,v+1)\t _{n}^{p+1}(u,v)\,,
\label{puv}
\end{eqnarray}

\begin{eqnarray}
\{p'nuv\}: \;\;\; &&
-\l _{p'u}\l _{vn}
\t _{n}^{p+1}(u,v+1)\t _{n}^{p}(u+1,v)
\!+\!\l _{p'v}\l _{un}
\t _{n}^{p+1}(u+1,v) \t _{n}^{p}(u,v+1)
\nonumber \\
&=&\l _{p'n}\l _{uv}
\t _{n-1}^{p+1}(u+1,v+1)\t _{n+1}^{p}(u,v)\,,
\label{p'nuv}
\end{eqnarray}

\vspace{0.2cm}

\begin{eqnarray}
\{p'uv\}: \;\;\;&&
\l _{p'v}\t _{n+1}^{p}(u,v+1)\t _{n}^{p+1}(u+1,v)
\!-\!\l _{p'u}\t _{n+1}^{p}(u+1,v)\t _{n}^{p+1}(u,v+1)
\nonumber \\
&=&\l _{uv}\t _{n}^{p+1}(u+1,v+1)\t _{n+1}^{p}(u,v)\,.
\label{p'uv}
\end{eqnarray}

\begin{eqnarray}
\{pnuv\}: \;\;\;&&
-\l _{pu}\l _{vn}
\t _{n+1}^{p}(u,v+1)\t _{n}^{p+1}(u+1,v)
\!+\!\l _{pv}\l _{un}
\t _{n+1}^{p}(u+1,v) \t _{n}^{p+1}(u,v+1)
\nonumber \\
&=&\l _{pn}\l _{uv}
\t _{n}^{p}(u+1,v+1)\t _{n+1}^{p+1}(u,v)\,,
\label{pnuv}
\end{eqnarray}

The above list of linearly independent
3-term bilinear equations is by no means complete.
The full list contains many other equations which
either follow from the already written ones or can be
derived from "higher" analogues of eqs.(\ref{tripl}),
(\ref{quadrupl}) for more than 4 variables
(which, in their turn, are algebraic corollaries of
eqs.\,(\ref{tripl})). Some of them are given in the
next subsection.

\subsection{Some useful corollaries of the basic equations}

Note that the pair of equations
(\ref{puv}), (\ref{p'nuv})
(respectively
(\ref{p'uv}), (\ref{pnuv}))
can be considered as a linear system for
$\t _{n}^{p+1}(u,v+1)\t _{n}^{p}(u+1,v)$,
$\t _{n}^{p+1}(u+1,v) \t _{n}^{p}(u,v+1)$
(respectively, for
$\t _{n+1}^{p}(u,v+1)\t _{n}^{p+1}(u+1,v)$,
$\t _{n+1}^{p}(u+1,v) \t _{n}^{p+1}(u,v+1)$).
Resolving these systems, we get

\begin{eqnarray}
&&\l _{p'v}\l _{un}
\t _{n}^{p}(u+1,v+1)\t _{n}^{p+1}(u,v)
+\l _{pu}\l _{p'n}
\t _{n-1}^{p+1}(u+1,v+1)\t _{n+1}^{p}(u,v)
\nonumber \\
&=&\Lambda
\t _{n}^{p+1}(u,v+1)\t _{n}^{p}(u+1,v)\,,
\label{uv3}
\end{eqnarray}

\begin{eqnarray}
&&\l _{p'u}\l _{vn}
\t _{n}^{p}(u+1,v+1)\t _{n}^{p+1}(u,v)
+\l _{pv}\l _{p'n}
\t _{n-1}^{p+1}(u+1,v+1)\t _{n+1}^{p}(u,v)
\nonumber \\
&=&\Lambda
\t _{n}^{p}(u,v+1)\t _{n}^{p+1}(u+1,v)\,,
\label{uv4}
\end{eqnarray}

\vspace{0.2cm}

\begin{eqnarray}
&&\l _{pv}\l _{un}
\t _{n}^{p+1}(u+1,v+1)\t _{n+1}^{p}(u,v)
+\l _{p'u}\l _{pn}
\t _{n}^{p}(u+1,v+1)\t _{n+1}^{p+1}(u,v)
\nonumber \\
&=&\Lambda
\t _{n+1}^{p}(u,v+1)\t _{n}^{p+1}(u+1,v)\,,
\label{uv5}
\end{eqnarray}

\begin{eqnarray}
&&\l _{pu}\l _{vn}
\t _{n}^{p+1}(u+1,v+1)\t _{n+1}^{p}(u,v)
+\l _{p'v}\l _{pn}
\t _{n}^{p}(u+1,v+1)\t _{n+1}^{p+1}(u,v)
\nonumber \\
&=&\Lambda
\t _{n}^{p+1}(u,v+1)\t _{n+1}^{p}(u+1,v)\,,
\label{uv6}
\end{eqnarray}

\noindent
where
\beq
\Lambda \equiv
\l _{pn}\l _{p'n}-\l _{un}\l _{vn}\,.
\label{Lambda}
\eeq
These equations are equally useful in what follows.
They are even more
informative than eqs.\,(\ref{puv})-(\ref{pnuv})
since allow one to "fuse" the variables, i.e. to put
$\l _{u}=\l _{v}$. (In (\ref{puv})-(\ref{pnuv}) this leads to
the identity $0=0$.) In this case $\t (u+1, v+1)$ converts into
$\t (u+2)$ and we get:

\begin{eqnarray}
&&\l _{un}\l _{p'u}
\t _{n}^{p+1}(u-1,v)\t _{n}^{p}(u+1,v)
+\l _{pu}\l _{p'n}
\t _{n+1}^{p}(u-1,v)\t _{n-1}^{p+1}(u+1,v)
\nonumber \\
&=&\big (\l _{p'n}\l _{pn}-\l _{un}^{2}\big )
\t _{n}^{p}(u,v)\t _{n}^{p+1}(u,v)\,,
\label{u3}
\end{eqnarray}

\begin{eqnarray}
&&\l _{un}\l _{pu}
\t _{n+1}^{p}(u-1,v)\t _{n}^{p+1}(u+1,v)
+\l _{pn}\l _{p'u}
\t _{n+1}^{p+1}(u-1,v)\t _{n}^{p}(u+1,v)
\nonumber \\
&=&\big (\l _{p'n}\l _{pn}-\l _{un}^{2}\big )
\t _{n+1}^{p}(u,v)\t _{n}^{p+1}(u,v)\,.
\label{u4}
\end{eqnarray}

\noindent
The same equations can be written down for $v$ in place of
$u$.

At last, we present two other useful
equations which are obtained
from (\ref{pp'u})-(\ref{pp'nu}),
(\ref{uv3})-(\ref{uv6}) and (\ref{u3}), (\ref{u4})
(together with their $v$-counterparts) by
a procedure similar to the one explained in the
remark after eq.\,(\ref{pp'nu}):

\begin{eqnarray}
&&\l _{pn}\l _{p'n}
\t _{n-1}^{p}(u+1,v+1)\t _{n+1}^{p}(u,v)
-\l _{un}\l _{vn}
\t _{n}^{p}(u+1,v+1)\t _{n}^{p}(u,v)
\nonumber \\
&=&\Lambda
\t _{n}^{p}(u+1,v)\t _{n}^{p}(u,v+1)\,,
\label{uv1}
\end{eqnarray}

\begin{eqnarray}
&&\l _{p'u}\l _{p'v}\l _{pn}
\t _{n}^{p-1}(u+1,v+1)\t _{n}^{p+1}(u,v)
-\l _{pu}\l _{pv}\l _{p'n}
\t _{n-1}^{p+1}(u+1,v+1)\t _{n+1}^{p-1}(u,v)
\nonumber \\
&=&\l _{p'p}\Lambda
\t _{n}^{p}(u+1,v)\t _{n}^{p}(u,v+1)\,.
\label{uv2}
\end{eqnarray}

To conclude the section, let us mention that
the complete list of linearly independent equations
for the $\t$-function (even restricted by the 3-term ones)
is somewhat longer.
Here we have selected those used in the sequel.

\section{Linearization of the discrete HM model}

The bilinear equations from the previous section
can be represented as compatibility conditions for
an overdetermined system of linear problems for
a "wave function" $\Psi$. This is what we mean by
the linearization. For us this is a systematic way
to find $L$-$M$-pairs.

\subsection{Scalar linear problems}

The bilinear equations (\ref{pp'n})-(\ref{pnuv}) follow from
compatibility of a system of {\it linear} equations for
a "wave function" $\Psi =\Psi _{n}^{p,p'}(u,v)$.
The prototype of linear equations for $\Psi$
is \cite{Hirota},\,\cite{DJM},\,\cite{SS}
(see also \cite{Zab} for a review):
\beq
\Psi (a+1,b)=\Psi (a, b+1)-\l _{ab}
\frac{\t (a,b)\t (a+1, b+1)}
{\t (a+1,b)\t (a, b+1)}
\Psi (a,b)\,,
\label{s2}
\eeq
where $a,b$ stand for any two elementary discrete variables.

The reduction condition
(\ref{b1}) for the $\Psi$-function reads
\beq
\Psi _{n}^{p+1, p'+1}(u,v)=z^2
\Psi _{n}^{p, p'}(u,v)\,,
\label{s1}
\eeq
where $z$ is a spectral parameter.
Hiding the variable $p'$ with the help of
this condition, we have:

\beq
\{nu\}: \;\;\;\;
\Psi _{n}^{p}(u+1)= \Psi _{n+1}^{p}(u)
-\l _{un}\frac{\t _{n}^{p}(u)\t _{n+1}^{p}(u+1)}
{\t _{n+1}^{p}(u)\t _{n}^{p}(u+1)}
\Psi _{n}^{p}(u)\,.
\label{nu}
\eeq

\beq
\{pu\}: \;\;\;\;
\Psi _{n}^{p+1}(u)= \Psi _{n}^{p}(u+1)
-\l _{pu}\frac{\t _{n}^{p}(u)\t _{n}^{p+1}(u+1)}
{\t _{n}^{p+1}(u)\t _{n}^{p}(u+1)}
\Psi _{n}^{p}(u)\,,
\label{pu}
\eeq

\beq
\{p'u\}: \;\;\;\;
z^2 \Psi _{n+1}^{p}(u)= \Psi _{n}^{p+1}(u+1)
-\l _{p'u}\frac{\t _{n}^{p+1}(u)\t _{n+1}^{p}(u+1)}
{\t _{n+1}^{p}(u)\t _{n}^{p+1}(u+1)}
\Psi _{n}^{p+1}(u)
\label{p'u}
\eeq

\noindent
(and similar equations for $v$ in place of $u$).
In (\ref{nu})-(\ref{p'u}),
$v$ is supposed to be the same everywhere and, therefore,
skipped.
These equations are basic tools for deriving
matrix $L$-$M$-pairs.

\subsection{Vector linear problem}

Combining equations (\ref{nu})-(\ref{p'u}), one can represent
translation of the vector
$
\left ( \begin{array}{c}
\Psi _{n}^{p}(u)\\
\Psi _{n}^{p+1}(u) \end{array}\right )
$
along the $u$-direction in the matrix form:
\beq
\left (\! \begin{array}{c}
\Psi _{n}^{p}(u\!+\!1)\\ \\ \\ \\
\Psi _{n}^{p+1}(u\!+\!1) \end{array}\! \right )
\!=\!\left (\! \begin{array}{ccc}
\!\l _{pu}
\displaystyle{\frac{\t _{n}^{p}(u)\t _{n}^{p+1}(u)}
{\t _{n}^{p+1}(u)\t _{n}^{p}(u\!+\!1)}}
&&1 \\ && \\
\!z^2 \l _{pn}
\displaystyle{\frac{\t _{n}^{p}(u)\t _{n+1}^{p+1}(u)}
{\t _{n}^{p+1}(u)\t _{n+1}^{p}(u)}}
&&
\!z^2 \!+\!\l _{p'u}
\displaystyle{\frac{\t _{n}^{p+1}(u)\t _{n+1}^{p}(u\!+\!1)}
{\t _{n+1}^{p}(u)\t _{n}^{p+1}(u\!+\!1)}}
\end{array}\!\right )\!
\left (\! \begin{array}{c}
\Psi _{n}^{p}(u)\\ \\ \\  \\
\Psi _{n}^{p+1}(u) \end{array}\! \right ).
\label{m1}
\eeq

Let us change the gauge passing to the wave function
\beq
\left ( \begin{array}{c}
\Phi _{1}(u)\\ \\
\Phi _{2}(u) \end{array} \right )
= {\cal D}(u)
\left ( \begin{array}{c}
\Psi _{n}^{p}(u)\\ \\
\Psi _{n}^{p+1}(u) \end{array} \right )
\label{m2}
\eeq
where ${\cal D}(u)$ is the diagonal matrix
$$
{\cal D}(u)=\left (\t _{n+1}^{p}(u)
\t _{n}^{p+1}(u)\right )^{-\frac{1}{2}}
\left ( \begin{array}{ccc}
\l _{pn}^{\frac{1}{4}}z^{\frac{1}{2}}\t _{n}^{p}(u)
&&0 \\ && \\
0&& \l _{pn}^{-\frac{1}{4}}z^{-\frac{1}{2}}
\t _{n}^{p+1}(u)\end{array}
\right ).
$$
Then the linear problem (\ref{m1}) acquires the form
\beq
\left ( \begin{array}{c}
\Phi _{1}(u+1)\\ \\
\Phi _{2}(u+1) \end{array} \right )
= L^{(u)}(z)
\left ( \begin{array}{c}
\Phi _{1}(u)\\ \\
\Phi _{2}(u) \end{array} \right )\,.
\label{m4}
\eeq
The $L$-operator $L^{(u)}(z)$  can be compactly written
in terms of the three fields

\beq
\psi ^{0}(u,v)=\frac{\t _{n}^{p+1}(u,v)}
{\t _{n+1}^{p}(u,v)}\,,
\;\;\;\;\;\;\;\;
\psi ^{-}(u,v)=\frac{\t _{n}^{p}(u,v)}
{\t _{n+1}^{p}(u,v)}\,,
\;\;\;\;\;\;\;\;
\psi ^{+}(u,v)=\frac{\t _{n+1}^{p+1}(u,v)}
{\t _{n}^{p+1}(u,v)}\,.
\label{psi}
\eeq

\noindent
For brevity we also use the notation
$\phi (u,v)\equiv \left [ \psi ^{0}(u,v)\right ]^{\frac{1}{2}}.$
The $L$-operator reads
\beq
L^{(u)}(z)=\left (
\begin{array}{ccc}
\l _{pu}\displaystyle{\frac{\phi (u+1)}{\phi (u)}}
&&z\l _{pn}^{\frac{1}{2}}
\displaystyle{\frac{\psi ^{-}(u+1)}{\phi (u)\phi (u+1)}} \\
&& \\z\l _{pn}^{\frac{1}{2}}\phi (u)\phi (u+1)\psi ^{+}(u) &&
\l _{p'u}\displaystyle{\frac{\phi (u)}{\phi (u+1)}}+z^2
\displaystyle{\frac{\phi (u+1)}{\phi (u)}}
\end{array}\right ).
\label{m5}
\eeq
Note that
\beq
\mbox{det} L^{(u)}(z)=\l _{pu}\l _{p'u}-\l _{un}z^2\,.
\label{m6}
\eeq
We call the $L^{(u)}(z)$ {\it chiral $L$-operator}
because it shifts the chiral variable of the wave function
via (\ref{m4}).
In the next subsection we study the discrete zero curvature
condition with matrices of the type (\ref{m5}).

\subsection{The zero curvature condition for
chiral $L$-operators}

Let

\vspace{0.1cm}

$$
\begin{array}{ccccc}
&\left |\phantom{\frac{A^A}{A^A}}\right. & &
\left |\phantom{\frac{A^A}{A^A}}\right. & \\
\frac{\phantom{aaaaaaa}}{\phantom{aaaaaaa}}\!\!\!&
\!\!\scriptstyle{C=(u,v+1)}\!\!&
\!\!\!\frac{\phantom{aaaaaaa}}{\phantom{aaaaaaa}}\!\!\!
&\!\!\scriptstyle{D=(u+1, v+1)}\!\!&
\!\!\!\frac{\phantom{aaaaaaa}}{\phantom{aaaaaaa}}\\
&\left |\phantom{\frac{A^A}{A^A}}\right. & &
\left |\phantom{\frac{A^A}{A^A}}\right. & \\
\frac{\phantom{aaaaaaa}}{\phantom{aaaaaaa}}\!\!\!&
\!\!\scriptstyle{A=(u,v)}\!\!&
\!\!\!\frac{\phantom{aaaaaaa}}{\phantom{aaaaaaa}}\!\!\!
&\!\!\scriptstyle{B=(u+1, v)}\!\!&
\!\!\!\frac{\phantom{aaaaaaa}}{\phantom{aaaaaaa}}\\
&\left |\phantom{\frac{A^A}{A^A}}\right. & &
\left |\phantom{\frac{A^A}{A^A}}\right. & \\
\end{array}
$$

\vspace{0.3cm}

\noindent
be an elementary cell of the $u,v$-lattice\footnote{In general
the coordinate axes are not orthogonal to each other;
in particular, at $\l _{u}=\l _{v}$ the lattice collapses
to a 1D one.}.
In this notation (borrowed from \cite{FadVol}-\cite{Fad})
the $L$-operator (\ref{m5}) reads:

\beq
L^{(u)}_{B\leftarrow A}(z)=\left (
\begin{array}{ccc}
\l _{pu}\displaystyle{\frac{\phi (B)}{\phi (A)}}
&&z\l _{pn}^{\frac{1}{2}}
\displaystyle{\frac{\psi ^{-}(B)}
{\phi (A)\phi (B)}} \\ && \\
z\l _{pn}^{\frac{1}{2}}\phi (A)\phi (B)\psi ^{+}(A) &&
\l _{p'u}\displaystyle{\frac{\phi (A)}{\phi (B)}}+z^2
\displaystyle{\frac{\phi (B)}{\phi (A)}}
\end{array}\right ).
\label{zc1}
\eeq

\noindent
Similarly, we introduce another chiral $L$-operator,
$L^{(v)}_{C\leftarrow A}(z)$, which is given by the same
formula with $\phi (C)$ in place of
$\phi (B)$ and $\l _{pv}$,  $\l _{p'v}$ in place of
$\l _{pu}$,  $\l _{p'u}$, respectively.

The discrete zero curvature condition
\beq
L^{(v)}_{D\leftarrow B}(z) L^{(u)}_{B\leftarrow A}(z)
=L^{(u)}_{D\leftarrow C}(z) L^{(v)}_{C\leftarrow A}(z)
\label{zc2}
\eeq
is equivalent to the following
non-linear equations of motion for the fields
$\psi ^{0}$, $\psi ^{\pm}$:

\begin{eqnarray}
&&\big (\l _{p'u}\psi ^{0}(C)- \l _{p'v}\psi ^{0}(B)\big )
\big (\psi ^{0}(A)\psi ^{0}(A)
-\psi ^{0}(B)\psi ^{0}(C)\big )
\nonumber \\
&=&\l _{pn}\psi ^{0}(B)\psi ^{0}(C)\psi ^{0}(D)
\big [\psi ^{+}(C)\psi ^{-}(C)-
\psi ^{+}(B)\psi ^{-}(B)\big ]\,,
\label{zc3}
\end{eqnarray}

\begin{eqnarray}
&&\l _{pv}\psi ^{0}(D)\psi ^{0}(C)\psi ^{-}(B)
+\l _{p'u}\psi ^{0}(A)\psi ^{0}(C)\psi ^{-}(D)
\nonumber \\
&=&\l _{pu}\psi ^{0}(D)\psi ^{0}(B)\psi ^{-}(C)
+\l _{p'v}\psi ^{0}(A)\psi ^{0}(B)\psi ^{-}(D)\,,
\label{zc4}
\end{eqnarray}

\begin{eqnarray}
&&\l _{pv}\psi ^{0}(D)\psi ^{0}(C)\psi ^{+}(C)
+\l _{p'u}\psi ^{0}(A)\psi ^{0}(C)\psi ^{+}(A)
\nonumber \\
&=&\l _{pu}\psi ^{0}(D)\psi ^{0}(B)\psi ^{+}(B)
+\l _{p'v}\psi ^{0}(A)\psi ^{0}(B)\psi ^{+}(A)\,.
\label{zc5}
\end{eqnarray}

\noindent
Equations of motion for the discrete HM model of $XXX$ type
given in \cite{DJM} are written for
a different choice of dynamical variables. It would be
useful to establish a direct correspondence between them.

\noindent
{\bf Remark}\,\,The discrete KdV equation
in the Faddeev-Volkov form \cite{FadVol},\,\cite{Volkov}
is reproduced from (\ref{zc4}) or (\ref{zc5})
in the limit $\l _{n}\rightarrow \infty$
when the $n$-dependence disappears, so that $\psi ^- =\psi ^+ =1$
and we are left with the single field $\psi ^{0}$.
The Faddeev-Volkov chiral
$L$-operators for the discrete KdV equation
are reproduced from (\ref{zc1}) in the same limit
provided the renormalizad
spectral parameter $\zeta = z\l ^{\frac{1}{2}}_{pn}$ is finite.

\subsection{Antichiral $L$-operators}

Here we introduce another type of $L$-operators
which will be called {\it antichiral}.

Let us put $\l _{u}=\l _{n}$ in the chiral $L$-operator
(\ref{zc1}) and, correspondingly, identify $u$ with $n$.
In this way we get the operator which generates the
translation
$$
A =(u, v, n)
\; \longrightarrow  \; A^{\uparrow} =(u,v,n+1)
$$
in the 3D lattice spanned by $u,v,n$:

\begin{eqnarray}
&&L^{(n)}_{A^{\uparrow}\leftarrow A}(z)=\left (
\begin{array}{ccc}
\l _{pn}\displaystyle{\frac{\phi (A^{\uparrow})}{\phi (A)}}
&&z\l _{pn}^{\frac{1}{2}}
\displaystyle{\frac{\psi ^{-}(A^{\uparrow})}
{\phi (A)\phi (A^{\uparrow})}}\\
&& \\
z\l _{pn}^{\frac{1}{2}}\phi (A)
\phi (A^{\uparrow})\psi ^{+}(A) &&
\l _{p'n}\displaystyle{\frac{\phi (A)}
{\phi (A^{\uparrow})}}+z^2
\displaystyle{\frac{\phi (A^{\uparrow})}{\phi (A)}}
\end{array}\right )
\nonumber \\
&& \nonumber \\
&=&\left (
\frac{\t _{n+1}^{p}(A)\t _{n+1}^{p+1}(A)}
{\t _{n}^{p+1}(A)\t _{n+2}^{p}(A)} \right )^{\frac{1}{2}}
\left ( \begin{array}{ccc}
\l _{pn} && z\l _{pn}^{\frac{1}{2}}
\displaystyle{\frac{\t _{n+1}^{p}(A)}{\t _{n+1}^{p+1}(A)}}
\\ && \\
z\l _{pn}^{\frac{1}{2}}
\displaystyle{\frac{\t _{n+1}^{p+1}(A)}{\t _{n+1}^{p}(A)}} &&
z^2 +\l _{p'n}
\displaystyle{\frac{\t _{n}^{p+1}(A)\t _{n+2}^{p}(A)}
{\t _{n+1}^{p}(A)\t _{n+1}^{p+1}(A)}}
\end{array} \right ).
\label{Ln}
\end{eqnarray}

\noindent
The similar notation
($B^{\uparrow} =(u+1, v, n+1)$,
$B^{\downarrow} =(u+1, v, n-1)$ etc)
will be used for the vertices in the upper and
lower layers of the 3D lattice.

Now we define the antichiral $L$-operator to be
\beq
\bar L _{C^{\downarrow}\leftarrow A}^{(\bar v )} (z)
=-\l _{pn}\l _{p'n} \left [
L _{C \leftarrow C^{\downarrow}}^{(n)}(z) \right ] ^{-1}
L _{C\leftarrow A}^{(v)} (z)\,.
\label{antichiral}
\eeq
It generates a discrete flow $\bar v$
(which we call {\it antichiral}) defined by the vector field
$\partial _{\bar v}\equiv -\partial _{n}+\partial _{v}$
in the space of variables. We can
display the flow $\bar v$ on the graph of
flows from Sect.\,2.2:

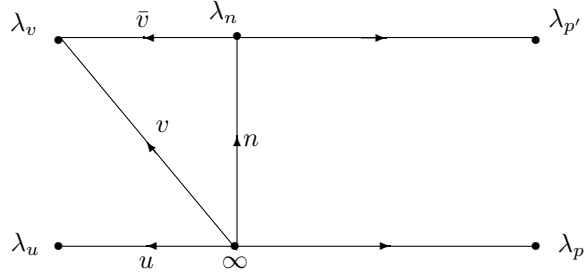
\begin{center}
\special{em:linewidth 0.4pt}
\unitlength 0.8mm
\linethickness{0.4pt}
\begin{picture}(95.00,59.00)
\emline{10.33}{20.00}{1}{90.00}{20.00}{2}
\put(90.00,20.00){\circle*{1.33}}
\put(10.67,20.00){\circle*{1.33}}
\put(40.00,20.00){\circle*{1.33}}
\emline{40.33}{20.00}{3}{40.33}{55.00}{4}
\put(40.33,55.00){\circle*{1.33}}
\put(40.33,38.14){\vector(0,1){0.2}}
\emline{40.33}{37.31}{5}{40.33}{38.14}{6}
\put(65.65,20.01){\vector(1,0){0.2}}
\emline{64.44}{20.01}{7}{65.65}{20.01}{8}
\put(25.48,20.01){\vector(-1,0){0.2}}
\emline{26.13}{20.01}{9}{25.48}{20.01}{10}
\put(7.67,20.33){\makebox(0,0)[rc]{$\lambda _{u}$}}
\put(40.67,59.00){\makebox(0,0)[rc]{$\lambda _{n}$}}
\put(93.67,20.00){\makebox(0,0)[lc]{$\lambda _{p}$}}
\put(40.00,17.33){\makebox(0,0)[cc]{$\infty$}}
\put(25.33,17.33){\makebox(0,0)[cc]{$u$}}
\put(42.67,37.67){\makebox(0,0)[cc]{$n$}}
\emline{40.33}{54.67}{11}{10.67}{54.67}{12}
\emline{11.00}{54.67}{13}{40.33}{19.33}{14}
\emline{40.00}{54.67}{15}{90.00}{54.67}{16}
\put(90.00,54.33){\circle*{1.33}}
\put(10.67,54.33){\circle*{1.33}}
\put(64.97,54.64){\vector(1,0){0.2}}
\emline{60.98}{54.64}{17}{64.97}{54.64}{18}
\put(25.54,37.19){\vector(-3,4){0.2}}
\emline{26.94}{35.44}{19}{25.54}{37.19}{20}
\put(25.02,54.67){\vector(-1,0){0.2}}
\emline{26.20}{54.67}{21}{25.02}{54.67}{22}
\put(5.00,57.00){\makebox(0,0)[cc]{$\lambda _{v}$}}
\put(95.00,57.33){\makebox(0,0)[cc]{$\lambda _{p'}$}}
\put(24.67,57.67){\makebox(0,0)[cc]{$\bar v$}}
\put(28.00,40.00){\makebox(0,0)[cc]{$v$}}
\end{picture}
\end{center}

\vspace{-1.0cm}

\noindent
Computing the r.h.s. of (\ref{antichiral}), we get:

\beq
\bar L^{(\bar v)}_{C^{\downarrow}\leftarrow A}(z)=\left (
\begin{array}{ccc}
\l _{vn}z^2
\displaystyle{\frac{\phi (A)}{\phi (C^{\downarrow})}}
-\l _{p'n}\l _{pv}
\displaystyle{\frac{\phi (C^{\downarrow})}{\phi (A)}}
&&-z\l _{pn}^{\frac{1}{2}}
\displaystyle{\frac{\psi ^{-}(A)}
{\phi (A)\phi (C^{\downarrow})}}
\\ && \\
-z\l _{pn}^{\frac{1}{2}}\phi (A)\phi (C^{\downarrow})
\psi ^{+}(C^{\downarrow})&&
-\l _{pn}\l _{p'v}\displaystyle{\frac{\phi (A)}
{\phi (C^{\downarrow})}}
\end{array}\right )
\label{antichiral1}
\eeq

\noindent
($v$-counterparts of
eqs.\,(\ref{pnu}), (\ref{p'nu}) have been used).

\noindent
{\bf Remark}\,\,Although we do not
need the antichiral flows themselves, we found it
instructive to mention them here in order to motivate
the choice of the $L$-operator in the next section.
Let us stress that
antichiral flows have "equal rights" with the chiral ones.
The symmetry can be seen from the figure and
from the explicit form of the $L$-operators. Note also
that in the case of the discrete KdV or SG models
there is no difference between chiral and
antichiral flows.

\section{"Composite" $L$ and $M$ operators}

The chiral and antichiral
$L$-operators introduced in the previous
section are building blocks for more complicated ones
obtained as their ordered products.
Not all of them have an $R$-matrix representation
of the desired form. Our next task is, therefore, to find
a special pair of flows $l,m$ such that
the corresponding $L$ and $M$ operators
meet the requirement. The lattice spanned by
$l,m$ will be called the {\it space-time lattice}.
It is embedded into the 3D lattice with coordinates
$u,v,n$.

Our experience in the lattice SG model suggests
that the good candidates are
"composite" $L$ and $M$ operators which generate
translations along
diagonals of the chiral space-time lattice.
In the discrete HM model this prescription
works literally for the $M$-operator
only. That is why, contrary to the tradition, we discuss
the $M$-operator first.
The $L$-operator requires an important
modification coming from the fact that in general
the space-time lattice is embedded into the 3D lattice
in a different way than the chiral one.

\subsection{The "composite" $M$-operator}

Consider the "composite" operator which transfers the
wave function along the diagonal $C\rightarrow B$:
\beq
\hat {\cal M}_{B\leftarrow C}(z)\equiv
z^{-1}
(\l _{vn}z^2 -\l _{pv}\l _{p'v})
L^{(u)}_{B\leftarrow A}(z)\left [
L^{(v)}_{C\leftarrow A}(z)\right ]^{-1}\,.
\label{c1}
\eeq
From (\ref{zc1}) we have:

\beq
\!\hat {\cal M}_{B\leftarrow C}(z)\!=\!
\left ( \!\!\begin{array}{ccc}
\!\!\!z\l _{un}\displaystyle{\frac{\phi (C)}{\phi (B)}}
\!-\!z^{-1}\l _{pu}\l _{p'v}\displaystyle{\frac{\phi (B)}{\phi (C)}}
&&
\!\l _{pn}^{\frac{1}{2}}
\displaystyle{\frac{\psi ^{-}(D)}{\phi (D)^{2}}}
\varphi _{{\cal M}}(C,B)
\\&&\\
\!\!\l _{pn}^{\frac{1}{2}}
\psi ^{+}(A)\phi (A)^{2}
\varphi _{{\cal M}}(C,B)
&&
\!\!z\l _{vn}\displaystyle{\frac{\phi (B)}{\phi (C)}}
\!-\!z^{-1}\!\l _{p'u}\l _{pv}\displaystyle{\frac{\phi (C)}{\phi (B)}}
\end{array} \!\! \right ),
\label{c2}
\eeq

$$
\varphi _{{\cal M}}(C,B)
=\l _{p'u}
\frac{\phi (C)}{\phi (B)}
-\l _{p'v}
\frac{\phi (B)}{\phi (C)}\,.
$$

\noindent
The diagonal elements are brought into this
form using eqs.\,(\ref{pnu}) and (\ref{p'nu}). The non-diagonal
elements are obtained with the help of
eqs.\,(\ref{zc4}), (\ref{zc5}).

Let us express the matrix elements in terms of the
$\t$-function. For any two vertices $X,Y$ of the lattice
we set
$$
W(X,Y)\equiv \left [\t _{n}^{p+1}(X)\t _{n+1}^{p}(X)
\t _{n}^{p+1}(Y)\t _{n+1}^{p}(Y) \right ]^{\frac{1}{2}}\,.
$$
Using eqs.\,(\ref{puv}), (\ref{p'uv}), we find:

\beq
\!\hat {\cal M}_{B\leftarrow C}(z)\!=\!\frac{1}{W(C,B)}
\!\left ( \!\!\begin{array}{ccc}
\!\begin{array}{c}\displaystyle{
\!z\l _{un}\t _{n}^{p+1}(C)\t _{n+1}^{p}(B)}\\
\displaystyle{
\!\!\!-z^{-1}\!\l _{pu}\l _{p'v}\t _{n}^{p+1}(B)\t _{n+1}^{p}(C)}
\end{array}
\!\!\!&&
\!\!-\l _{uv}\l _{pn}^{\frac{1}{2}}
\t _{n+1}^{p}(A)\t _{n}^{p}(D)
\\&&\\ &&\\
\!\!-\l _{uv}\l _{pn}^{\frac{1}{2}}
\t _{n+1}^{p+1}(A)\t _{n}^{p+1}(D)
&&
\!\!\begin{array}{c}\!\displaystyle{
\!\!z\l _{vn}\t _{n}^{p+1}(B)\t _{n+1}^{p}(C)}\\
\displaystyle{
\!\!-z^{-1}\!\l _{p'u}\l _{pv}\t _{n}^{p+1}(C)\t _{n+1}^{p}(B)}
\!\!\end{array}
\end{array} \!\!\!\right ).
\label{c3}
\eeq

Moreover, with the help of eqs.\,(\ref{uv5}), (\ref{uv6})
it is possible to express the matrix elements through the
$\t$-functions at the points $A$ and $D$ only, while the
dependence on the points $C$, $B$ is confined in the prefactor
$W^{-1}(C,B)$. Indeed, the r.h.s. of
eqs.\,(\ref{uv5}), (\ref{uv6}) are just the two terms on the
diagonal of the matrix in (\ref{c3}). Substituting them by
the left hand sides, we see that the matrix depends on $A$ and
$D$ only. This is used in Sect.\,5.

\subsection{The "composite" $L$-operator -- first version}

The first candidate is suggested by our experience in the
discrete SG model \cite{Zsg}:
\beq
{\cal L}^{(uv)}_{D\leftarrow A}(z)=z^{-1}L^{(v)}_{D\leftarrow B}(z)
L^{(u)}_{B\leftarrow A}(z)\,.
\label{c6}
\eeq
Using the bilinear equations from Sect.\,2 many times,
we can express its matrix elements through the
$\t$-function:
\beq
{\cal L}^{(uv)}_{D\leftarrow A}(z)=\frac{1}{\l _{uv}W(A,D)}
\left ( \begin{array}{ccc}
{\cal L}^{(uv)}_{11}&&
{\cal L}^{(uv)}_{12} \\&&\\
{\cal L}^{(uv)}_{21} &&
{\cal L}^{(uv)}_{22}
\end{array} \right )
\label{c7}
\eeq
with

\beq
{\cal L}^{(uv)}_{11}=
\l _{pv}\zeta _{u}(z)\t _{n+1}^{p}(B)\t _{n}^{p+1}(C)
-\l _{pu}\zeta _{v}(z)\t _{n+1}^{p}(C)\t _{n}^{p+1}(B)\,,
\label{c811}
\eeq

\beq
{\cal L}^{(uv)}_{12}=
\l _{pn}^{\frac{1}{2}}z\left (\zeta _{u}(z)
\t _{n+1}^{p}(B)\t _{n}^{p}(C)
-\zeta _{v}(z)\t _{n+1}^{p}(C)\t _{n}^{p}(B)
\phantom{A^{a^a}}\!\!\!\!\!\!\!\!\!\right ),
\label{c812}
\eeq

\beq
{\cal L}^{(uv)}_{21}=
\l _{pn}^{\frac{1}{2}}z\left (\zeta _{u}(z)
\t _{n+1}^{p+1}(B)\t _{n}^{p+1}(C)
-\zeta _{v}(z)\t _{n+1}^{p+1}(C)\t _{n}^{p+1}(B)\right ),
\label{c821}
\eeq

\begin{eqnarray}
&&{\cal L}^{(uv)}_{22}=
\frac{\zeta _{u}(z)\zeta _{v}(z)}{\l _{un}\l _{vn}}
\left (
\l _{p'v}\t _{n}^{p+1}(B)\t _{n+1}^{p}(C)
-\l _{p'u}\t _{n}^{p+1}(C)\t _{n+1}^{p}(B)\right )
\nonumber \\
&+&\frac{\l _{pn}}{\l _{un}\l _{vn}}\left (
\l _{p'v}\l _{un}\zeta _{u}(z)\t _{n+1}^{p+1}(B)\t _{n}^{p}(C)
-\l _{p'u}\l _{vn}\zeta _{v}(z)
\t _{n+1}^{p+1}(C)\t _{n}^{p}(B)\right ).
\label{c822}
\end{eqnarray}
Here
\beq
\zeta _{u}(z)=\l _{un}z^2 -\l _{pu}\l _{p'u}\,,
\;\;\;\;\;\;
\zeta _{v}(z)=\l _{vn}z^2 -\l _{pv}\l _{p'v}\,.
\label{c9}
\eeq

As it will be clear later,
the desired $R$-matrix representation
for this operator does not exist since the matrix elements
have a "wrong" dependence on $z$. We should seek for an
$L$-operator which would depend on $z$ as in (\ref{c3}).

\subsection{The "composite" $L$-operator -- improved version}

The above $L$-operator can be "improved" by an additional
translation in $n$.
The right choice is the operator which
generates the translation
$$A=(u,v,n)
\; \longrightarrow \; D^{\downarrow}=(u+1, v+1, n-1)$$
in the 3D lattice spanned by $u,v,n$. Its elementary cell
is shown in the figure:

\vspace{2mm}

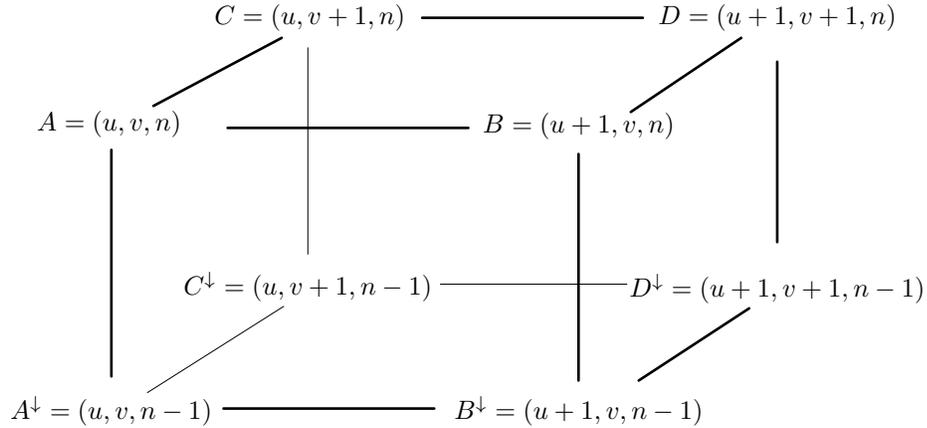
\begin{center}
\special{em:linewidth 0.4pt}
\unitlength 0.8mm
\linethickness{0.4pt}
\begin{picture}(125.27,85.33)
\put(-190.92,16.93){\circle*{0.00}}
\special{em:linewidth 1.0pt}
\emline{88.00}{25.00}{1}{88.00}{62.67}{2}
\emline{10.33}{63.33}{3}{10.33}{25.67}{4}
\emline{98.00}{25.00}{5}{116.33}{37.00}{6}
\emline{96.67}{69.67}{7}{115.00}{82.00}{8}
\emline{17.33}{70.67}{9}{38.67}{82.00}{10}
\emline{121.00}{78.00}{11}{121.00}{48.00}{12}
\put(10.00,67.67){\makebox(0,0)[cc]{$A=(u,v,n)$}}
\put(43.33,85.33){\makebox(0,0)[cc]{$C=(u,v+1,n)$}}
\put(88.00,67.33){\makebox(0,0)[cc]{$B=(u+1,v,n)$}}
\put(121.00,85.33){\makebox(0,0)[cc]{$D=(u+1, v+1, n)$}}
\put(10.33,20.00){\makebox(0,0)[cc]{$A^{\downarrow}=(u,v,n-1)$}}
\put(88.00,20.00){\makebox(0,0)[cc]{$B^{\downarrow}=(u+1, v,n-1)$}}
\put(121.00,40.33){\makebox(0,0)[cc]{$D^{\downarrow}=(u+1,v+1,n-1)$}}
\put(43.00,40.67){\makebox(0,0)[cc]{$C^{\downarrow}=(u,v+1,n-1)$}}
\emline{62.00}{85.33}{13}{98.67}{85.33}{14}
\emline{29.67}{67.00}{15}{69.67}{67.00}{16}
\emline{29.00}{20.33}{17}{64.00}{20.33}{18}
\put(-190.92,16.93){\circle*{0.00}}
\special{em:linewidth 0.05pt}
\emline{65.00}{41.00}{19}{96.00}{41.00}{20}
\emline{16.33}{23.00}{21}{39.00}{37.33}{22}
\emline{43.00}{80.33}{23}{43.00}{46.00}{24}
\end{picture}
\end{center}

\vspace{-1.0cm}

\noindent
One of possible ways to represent such an $L$-operator
in terms of already mentioned ones is:
\beq
\hat {\cal L}_{D^{\downarrow}\leftarrow A}(z)=
-\l _{pn}\l _{p'n}\left [ L^{(n)}_{D\leftarrow D^{\downarrow}}(z)
\right ]^{-1}
{\cal L}^{(uv)}_{D\leftarrow A}(z)
\label{k1}
\eeq
with $L^{(n)}_{D\leftarrow D^{\downarrow}}(z)$
as in (\ref{Ln}) and
${\cal L}_{D\leftarrow A}^{(uv)}(z)$ as in (\ref{c7}).
Matrix elements of
$\hat {\cal L}_{D^{\downarrow}\leftarrow A}(z)$ are
Laurent polynomials in $z$. From (\ref{k1}),
(\ref{zc1}) and (\ref{c811})-(\ref{c9}) we see that
the diagonal elements are of at most 3-d degree, the
right upper (resp., left lower) element is of at most
4-th (resp., 2-nd) degree.

It turns out that this $L$-operator actually has
a surprizingly simple structure -- much simpler than
one could expect from (\ref{k1}). In particular,
using the bilinear equations, one can show that the highest
degrees of $z$ cancel leaving us with at most $z$ and $z^{-1}$-terms
in the diagonal elements and $z$-independent terms in the
non-diagonal ones.

This cancellation could be expected from the following
equivalent representation of the $L$-operator
(\ref{k1}) which makes use of the antichiral
flow introduced in Sect.\,3.4:
$$
\hat {\cal L}_{D^{\downarrow}\leftarrow A}(z)=
L _{D^{\downarrow}\leftarrow C^{\downarrow}}^{(u)}(z)
\bar L _{C^{\downarrow}\leftarrow A}^{(\bar v )}(z)\,.
$$
This means that the flow generated by the $L$-operator
(\ref{k1}) is the "superposition" of the chiral and
antichiral ones.

Computing the $L$-operator by means of any one
of the two formulas, we find:

\beq
\!\!\!\!\hat {\cal L}_{D^{\downarrow}\leftarrow A}(z)\!=\!
\left ( \!\begin{array}{ccc}
\!\!z\l _{un}\l _{vn}\displaystyle{\frac{\phi (A)}
{\phi (D^{\downarrow})}}
\!+\!z^{-1}\!\l _{pu}\l _{pv}\l _{p'n}
\displaystyle{\frac{\phi (D^{\downarrow})}{\phi (A)}}
&&
\l _{pn}^{\frac{1}{2}}
\displaystyle{\frac{\psi ^{-}(B)}{\phi (B)^{2}}}
\varphi _{{\cal L}}(A, D^{\downarrow})
\\&&\\
\l _{pn}^{\frac{1}{2}} \psi ^{+}(C^{\downarrow})
\phi (C^{\downarrow})^{2}
\varphi _{{\cal L}}(A, D^{\downarrow})
&&
\!\!\!z\l _{pn}\l _{p'n}
\displaystyle{\frac{\phi (D^{\downarrow})}{\phi (A)}}
\!+\!z^{-1}\!\l _{pn}\l _{p'u}\l _{p'v}
\displaystyle{\frac{\phi (A)}{\phi (D^{\downarrow})}}
\end{array} \!\!\right ),
\label{k3}
\eeq

$$
\varphi _{{\cal L}}(A, D^{\downarrow})=
\l _{p'u}\l _{vn}
\displaystyle{\frac{\phi (A)}{\phi (D^{\downarrow})}}
+\l _{pv}\l _{p'n}
\displaystyle{\frac{\phi (D^{\downarrow})}{\phi (A)}}\,.
$$

\noindent
In terms of the $\t$-function we have:

\beq
\!\hat {\cal L}_{D^{\downarrow}\leftarrow A}(z)\!=\!
\frac{1}{W(A,D^{\downarrow})}
\left ( \!\! \begin{array}{ccc}
\!\!\begin{array}{c}\displaystyle{
\!z\l _{un}\l _{vn}\t _{n}^{p}(D)\t _{n}^{p+1}(A)}
\\
\displaystyle{
\!\!\!+z^{-1}\!\l _{pu}\l _{pv}\l _{p'n}
\t _{n-1}^{p+1}(D)\t _{n+1}^{p}(A)}\!
\end{array}
\!\!\!&&
\!\!\Lambda \l _{pn}^{\frac{1}{2}}
\t _{n}^{p}(B)\t _{n}^{p}(C)
\\&&\\ &&\\
\Lambda \l _{pn}^{\frac{1}{2}}
\t _{n}^{p+1}(B)\t _{n}^{p+1}(C)
&&
\!\!\begin{array}{c} \!\!\displaystyle{
\!\!\!z\l _{pn}\l _{p'n}\t _{n-1}^{p+1}(D)\t _{n+1}^{p}(A)}
\\
\displaystyle{
\!\!\!+z^{-1}\!\l _{pn}\l _{p'u}\l _{p'v}
\t _{n}^{p}(D)\t _{n}^{p+1}(A)}\!
\end{array}
\end{array} \!\!\!\right )
\label{k4}
\eeq

\noindent
($\Lambda$ is given in (\ref{Lambda})).

\subsection{The space-time lattice}

Let us consider
the 2D lattice spanned by coordinates $l,m$
which are flows generated by the
$L$ and $M$ introduced above.
The vector fields
$$
\partial _{l}\equiv \partial _{u}
+\partial _{v}-\partial _{n}\,,
\;\;\;\;\;\;\;\;
\partial _{m}\equiv \partial _{u}
-\partial _{v}
$$
are then "coordinate axes" of this lattice.
(Note the equivalent representations of these vector
fields in terms of antichiral flows:
$\partial _{l}= \partial _{u}+\partial _{\bar v}
=\partial _{\bar u}+\partial _{v}$,
$\partial _{m}= \partial _{\bar u}-\partial _{\bar v}$.)
Shifting the origin of coordinates,
we always can embed this lattice into the 3D one
as the 2D plane defined by the
homogeneous linear equation
\beq
u+v+2n=0
\label{plane}
\eeq
provided $(u,v,n)\in {\bf Z}^{3}$.
In the sequel this 2D lattice
is refered to as {\it the space-time lattice} (STL).
Note that the chiral lattice is the plane $n=0$, i.e.
it is the linear projection of the STL along the $n$-direction.

{\it The space-time coordinates} $l,m$ are introduced by
the formulas
\beq
l=\frac{1}{2}(u+v),\,  \;\;\;\;\;\;\;\;
m=\frac{1}{2}(u-v)
\label{lm}
\eeq
provided the point $(u,v,n)$ belongs to the STL. (Under the
same condition $l$ could be
equivalently defined as $l=-n$.)

\noindent
{\bf Remark}\,\,After an obvious matching of the notation,
the zero curvature condition
on the elementary cell of the STL
for the constructed "composite"
$L$-$M$-pair takes the form (\ref{dzc}). It gives rise to a system
of non-linear equations for fields at the vertices of
eight neighbouring cubs of the 3D lattice. This system is
a direct consequence of the basic equations of
motion (\ref{zc3})-(\ref{zc5}).

\section{$R$-matrix representation of the $L$-$M$-pair}

Our goal is to represent the $L$ and $M$ operators
(\ref{c3}), (\ref{k4})
constructed in the previous section
as convolutions of a quantum $R$-matrix
with some vectors in its "quantum" space.

\subsection{The quantum $R$-matrix}

Consider the following quantum $R$-matrix with the
spectral parameter $z$:

\begin{eqnarray}
R(z)&=&\frac{1}{4}\left [ 2a(z)+b_+(z)+b_-(z)\right ]
I\otimes I
+\frac{1}{4}\left [ 2a(z)-b_+(z) - b_-(z)\right ]
\sigma _{3}\otimes \sigma _{3}
\nonumber\\
&+&\frac{1}{4} \left [ b_+(z)-b_-(z)\right ]
\left ( I\otimes \sigma _{3}- \sigma _{3}\otimes I\right )
+\frac{1}{2} c(z)
\left ( \sigma _{1}\otimes \sigma _{1}
+\sigma _{2}\otimes \sigma _{2}\right )
\label{q1}
\end{eqnarray}
$$
=\,\,\left (
\begin{array}{ccccccc}
a(z)&&0&&0&&0 \\
&&&&&&\\
0&& b_{+}(z) && c(z) && 0 \\
&&&&&&\\
0 && c(z) && b_{-}(z) && 0 \\
&&&&&&\\
0&& 0 && 0&&  a(z)
\end{array} \right )
$$

\noindent
($\sigma_i$ are Pauli matrices), where
\beq
a(z)=qz-q^{-1}z^{-1}, \;\;\;\;\;
b_{\pm}(z)=\xi ^{\pm 1}(z-z^{-1}), \;\;\;\;\;
c(z)=q-q^{-1}.
\label{q2}
\eeq
When necessary, we write
$R(z)=R(z;q,\xi )$;
$q$ is the "quantum" parameter and
$\xi$ is the parameter of
Drinfeld's twist \cite{D},\,\cite{Resh}:
$$
R(z;q,\xi )=F(\xi)R(z;q,1)F(\xi)\,,
$$
$$
F(\xi)=
\xi ^{\frac{1}{4}(I\otimes \sigma _{3}- \sigma _{3}\otimes I)}=
\mbox{diag} \left (1,\xi^{\frac{1}{2}},
\xi^{-\frac{1}{2}},1\right )\,.
$$
The $R$-matrix (\ref{q1}) satisfies the
quantum Yang-Baxter equation
for any $q$, $\xi$:
\beq
R_{12}(z_1 /z_2)R_{13}(z_1 /z_3)
R_{23}(z_2 /z_3)=R_{23}(z_2 /z_3)
R_{13}(z_1 /z_3)R_{12}(z_1 /z_2)\,,
\label{qyb}
\eeq
where $R_{12}(z)$ acts in
${\bf C}^2\otimes {\bf C}^2\otimes {\bf C}^2$ as the $R(z)$ in the
first and the second spaces and as the $I$ in the third one
(similarly for $R_{13}(z)$ and $R_{23}(z)$).

The first (resp., second) space of tensor products
in (\ref{q1}) is called "quantum" (resp., "auxiliary") space.
It is convenient to represent
the $R$-matrix as a 2$\times$2 block matrix in the
"auxiliary" space. Let $i, i'$ number block rows and
columns and let $j, j'$ number rows and columns inside each block
(i.e., in the "quantum" space). Then matrix elements of the
$R$-matrix (\ref{q1}) are denoted by $R(z)^{ii'}_{jj'}$.

Let
$\big |\a \big > ,\,
\big |\b \big > $
be two vectors in the quantum space (see (\ref{r5})).
Each block of the
$R$-matrix is an operator in the quantum space. Consider
its action to $\big |\a \big >$ and subsequent scalar product with
$\big <\b \big |$.
The result
is a 2$\times$2 matrix in the auxiliary space:
$$
\big <\b \big | R(z) \big |\a \big >_{ii'}
=\sum _{jj'}R(z)^{ii'}_{jj'}\a _{j'}\b _{j}.
$$
Substituting the matrix (\ref{q1}), we find:

\begin{eqnarray}
\!\big <\b \big | R(z) \big |\a \big >\! &\!=\!&
\!\left (
\!\begin{array}{ccc}\!
\b _{1}\a _{1}a(z) +\b _{2}\a _{2}b_{+}(z) &&
\!\b _{2}\a _{1}c(z) \\ && \\
\!\b _{1}\a _{2}c(z) &&
\!\b _{1}\a _{1}b_{-}(z) +\b _{2}\a _{2}a(z)
\end{array}\!\! \right )
\nonumber \\
&& \nonumber \\
&=&\left ( \!\! \begin{array}{ccc}
\!\!\begin{array}{c}
\displaystyle{
\!z (q\b _{1}\a _{1}+\xi \b _{2}\a _{2})}
\\
\displaystyle{
-z^{-1}\! (q^{-1}\!\b _{1}\a _{1}+\xi \b _{2}\a _{2})}
\end{array}
\!\!&&
(q-q^{-1})\b _{2}\a _{1}
\\&&\\ &&\\
(q-q^{-1})\b _{1}\a _{2}
&&
\!\!\begin{array}{c}
\displaystyle{
\!z q \xi ^{-1}\!
(q^{-1}\!\b _{1}\a _{1}+\xi \b _{2}\a _{2})}
\\
\displaystyle{
-z^{-1}q^{-1}\xi ^{-1}
(q\b _{1}\a _{1}+\xi \b _{2}\a _{2})}
\end{array}
\end{array} \!\!\right ).
\label{q3}
\end{eqnarray}

\noindent
Hereafter all scalar products of the type
$\big <\b \big | R(z) \big |\a \big >$ are understood
as to be taken in the {\it first} ("quantum") space.

\subsection{The $L$-$M$-pair in terms of the $R$-matrix}

We are going to bring the $L$ and $M$ operators
(\ref{k3}), (\ref{c2}) into the form (\ref{q3}).
The best result is
achieved after the diagonal gauge transformation
\beq
\hat {\cal M}_{B\leftarrow C}(z)
\longrightarrow {\cal M}_{B\leftarrow C}(z)= \left (
\frac{\t _{n}^{p+1}(B)\t _{n+1}^{p}(B)}
{\t _{n}^{p+1}(C)\t _{n+1}^{p}(C)}
\right )^{\frac{1}{2}} \hat {\cal M}_{B\leftarrow C}(z)\,,
\label{q4a}
\eeq
\beq
\hat {\cal L}_{D^{\downarrow}\leftarrow A}(z)
\longrightarrow
{\cal L}_{D^{\downarrow}\leftarrow A}(z)=
\left ( \frac{\t _{n-1}^{p+1}(D)\t _{n}^{p}(D)}
{\t _{n}^{p+1}(A)\t _{n+1}^{p}(A)} \right )^{\frac{1}{2}}
\hat {\cal L}_{D^{\downarrow}\leftarrow A}(z)
\label{q4b}
\eeq
and shift of the spectral parameter
\beq
z\rightarrow kz\,, \;\;\;\;\;\;\;\;
k = \left (\frac{\l _{pu}\l _{p'u}}
{\l _{un}}\right )^{\frac{1}{2}}.
\label{q5}
\eeq
The matrices
${\cal M}_{B\leftarrow C}(kz)$,
${\cal L}_{D^{\downarrow}\leftarrow A}(kz)$
are to be expressed through the $\t$-function
as is explained at the end of Sect.\,4.1.

In order to present the result, let us
introduce the vectors

\beq
\big |\a (u,v,n)\big >=
\left ( \begin{array}{c} \mu\tau _{n}^{p}(u,v) \\ \\
\mu ^{-1} \tau _{n}^{p+1}(u,v)
\end{array} \right ), \;\;\;\;\;\;\;\;
\big |\b (u,v,n) \big > =
\left ( \begin{array}{c} \mu\tau _{n}^{p+1}(u,v) \\ \\
\mu ^{-1} \tau _{n}^{p}(u,v)
\end{array} \right ),
\label{vect}
\eeq

\noindent
where
\beq
\mu =\left (\frac{\l _{un}\l _{pu}}
{\l _{pn}\l _{p'u}}\right )^{\frac{1}{4}}
\label{mu}
\eeq

\noindent
and identify the parameters as follows:

\beq
q=\left (\frac{\l _{vn}\l _{pu}\l _{p'u}}
{\l _{un}\l _{pv}\l _{p'v}}
\right )^{\frac{1}{2}},
\;\;\;\;\;\;\;
\xi =\left (\frac{\l _{un}\l _{pu}\l _{p'v}}
{\l _{vn}\l _{pv}\l _{p'u}}
\right )^{\frac{1}{2}},
\;\;\;\;\;\;\;
\xi ' = -\left ( \frac{\l _{un}\l _{vn} \l _{pu}\l _{pv}}
{\l _{pn}^{2} \l _{p'u}\l _{p'v}}\right )^{\frac{1}{2}}
\label{q6}
\eeq

\noindent
(note that $\xi \xi ' =-\mu ^{4}$).
For brevity we also set
$$
\rho \equiv \left [\l _{pn}\l _{un}\l _{vn}
\l _{pu}\l _{pv}\l _{p'u}\l _{p'v}\right ]^{\frac{1}{2}}\,,
$$
then $q-q^{-1}=-\l _{pn}^{\frac{1}{2}}
\l _{uv}\Lambda \rho ^{-1}$
(recall that $\Lambda$ is given by (\ref{Lambda})).

After these preparations, the comparison with (\ref{q3})
immediately yields the formulas

\beq
\big <\b (B)\big | R(z;q, \xi ') \big |\a (C)\big >
=-\l _{uv}\rho ^{-1}
\t _{n}^{p+1}(A) \t _{n+1}^{p}(A)
{\cal L}_{D^{\downarrow}\leftarrow A}(kz)\,,
\label{q8}
\eeq

\beq
\big <\b (D)\big | R(z;q, \xi ) \big |\a (A^{\uparrow})\big >
=\Lambda \rho ^{-1}
\t _{n}^{p+1}(C) \t _{n+1}^{p}(C)
{\cal M}_{B\leftarrow C}(kz)\,.
\label{q7}
\eeq

\noindent
To recast them into the form (\ref{r9}), (\ref{r8})
respectively, we, first of all,
fix the constant time slice $m=0$ of the STL. It is
embedded into the 3D lattice as the 1D lattice with vertices
at the points $A_l =(l,\, l,-l)$, $l\in {\bf Z}$. Let us introduce
the similar notation for other points:
$B_l =(l+1,l,-l)$,
$\bar B_l =(l+1,l-1,-l)$,
$D_l^{\downarrow} =A_{l+1}=(l+1,l+1,-l-1)$ etc
and put
\beq
{\cal L}_{l}(z)=
{\cal L}_{D_{l}^{\downarrow}\leftarrow A_l }(z)\,,
\;\;\;\;\;\;\;\;
{\cal M}_{l}(z)=
{\cal M}_{\bar B_{l} \leftarrow A_l }(z)\,.
\label{q9}
\eeq
Now we notice that
\beq
\t _{n+1}^{p}(u,v) \t _{n}^{p+1}(u,v)=
\frac{(\l _{pn} \l _{un} \l _{pu} \l _{p'u})^{\frac{1}{2}}}
{\l _{pn} \l _{p'n}- \l _{un}^{2} }\,
\big <\b (u+1,v,n)\big | \a (u-1,v,n+1)\big >
\label{q14}
\eeq
due to eq.\,(\ref{u3}).
This relation allows us to rewrite
eqs.\,(\ref{q8}), (\ref{q7}) in the form

\beq
{\cal L}_{l}(kz)=
-\frac{\gamma}{\l _{uv}}
\frac{
\big <\b (B_l )\big | R(z;q, \xi ') \big |\a (C_l )\big >}
{\big <\b (B_l )\big | \a (C_{l-1} )\big >}\,,
\label{q12}
\eeq

\beq
{\cal M}_{l}(kz)=
\frac{\gamma}{\Lambda}
\frac{
\big <\b (B_l )\big | R(z;q, \xi )
\big |\check \b (B_{l-1} )\big >}
{\big <\b (B_l )\big | \a (C_{l-1} )\big >}\,,
\label{q13}
\eeq

\noindent
where
$$
\big |\check \b (B_{l} )\big > =
\left ( \begin{array}{c}
\mu \t _{-l}^{p}(l+1,l) \\ \\
\mu ^{-1}\,\t _{-l}^{p+1}(l+1,l) \end{array}\right )
=\left ( \begin{array}{ccc}0&&\mu ^{2}\\ \\
\mu ^{-2}&&0 \end{array}\right )
\big |\b (B_{l} )\big >
$$
and $\gamma$ is a constant:
$$
\gamma =\left [ \l _{vn} \l _{pv} \l _{p'v} \right ]^{\frac{1}{2}}
\left ( \l _{pn} \l _{p'n} -\l _{un}^{2} \right )\,.
$$
These formulas differ from the ones
announced in the Introduction merely by the irrelevant
prefactors. To match the notation,
let us rename the vectors:
$|\a (C_l )\big > \rightarrow |\a (l)\big >$,
$|\b (B_l )\big > \rightarrow |\b (l)\big >$,
so from now on we set
\beq
\big |\a (l)\big >=
\left ( \begin{array}{c}
\mu \t _{-l}^{p}(l,l+1) \\ \\ \mu ^{-1}\,
\t _{-l}^{p+1}(l,l+1) \end{array}\right )\,,
\;\;\;\;\;\;\;\;
\big |\b (l)\big >=
\left ( \begin{array}{c}
\mu \t _{-l}^{p+1}(l+1,l) \\ \\
\mu ^{-1}\,\t _{-l}^{p}(l+1,l) \end{array}\right )\,.
\label{q11}
\eeq
Their location
in the 3D lattice is shown in the figure:

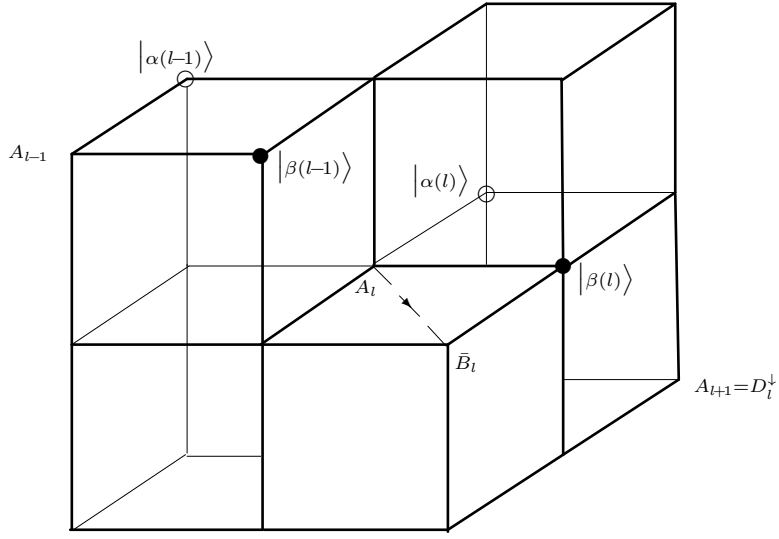
\begin{center}
\special{em:linewidth 0.4pt}
\unitlength 1.00mm
\linethickness{0.4pt}
\begin{picture}(92.44,80.00)
\put(-190.92,16.93){\circle*{0.00}}
\special{em:linewidth 1.00pt}
\emline{9.56}{10.00}{1}{35.11}{10.00}{2}
\emline{35.11}{10.00}{3}{35.11}{34.67}{4}
\emline{35.11}{34.67}{5}{9.78}{34.67}{6}
\emline{9.78}{34.67}{7}{9.78}{10.00}{8}
\emline{35.11}{34.67}{9}{59.78}{34.67}{10}
\emline{59.78}{34.67}{11}{59.78}{9.78}{12}
\emline{35.33}{10.00}{13}{59.78}{10.00}{14}
\emline{9.78}{34.67}{15}{9.78}{60.00}{16}
\emline{9.78}{60.00}{17}{35.11}{60.00}{18}
\emline{35.11}{60.00}{19}{35.11}{34.67}{20}
\emline{9.78}{60.00}{21}{25.11}{70.00}{22}
\emline{25.11}{70.00}{23}{49.78}{70.00}{24}
\emline{49.78}{70.00}{25}{35.33}{60.00}{26}
\emline{59.78}{34.67}{27}{75.11}{45.11}{28}
\emline{75.11}{45.11}{29}{75.11}{20.00}{30}
\emline{75.11}{20.00}{31}{59.78}{10.00}{32}
\emline{34.89}{34.67}{33}{50.00}{45.11}{34}
\emline{50.00}{45.11}{35}{50.00}{70.00}{36}
\emline{50.22}{45.11}{37}{74.89}{45.11}{38}
\emline{50.00}{70.00}{39}{74.89}{70.00}{40}
\emline{74.89}{70.00}{41}{75.11}{44.89}{42}
\emline{50.00}{70.22}{43}{64.89}{80.00}{44}
\emline{64.89}{80.00}{45}{90.00}{80.00}{46}
\emline{90.00}{80.00}{47}{74.89}{69.78}{48}
\emline{75.33}{44.89}{49}{90.00}{54.89}{50}
\emline{90.00}{54.89}{51}{90.00}{79.78}{52}
\emline{90.00}{54.67}{53}{90.44}{30.00}{54}
\emline{90.44}{30.00}{55}{75.11}{20.00}{56}
\put(4.00,60.00){\makebox(0,0)[cc]{$\scriptstyle{A_{l\!-\!1}}$}}
\put(48.89,42.22){\makebox(0,0)[cc]{$\scriptstyle{A_{l}}$}}
\put(62.22,32.22){\makebox(0,0)[cc]{$
\scriptstyle{\bar B_{l}}$}}
\put(92.44,29.11){\makebox(0,0)[lc]{$
\scriptstyle{A_{l\!+\!1}=D_{l}^{\downarrow}}$}}
\put(23.56,72.89){\makebox(0,0)[cc]{$
\scriptstyle{\big |\alpha (l\!-\!1)\big >}$}}
\put(36.89,58.22){\makebox(0,0)[lc]{$
\scriptstyle{\big |\beta (l\!-\!1)\big >}$}}
\put(76.89,43.33){\makebox(0,0)[lc]{$
\scriptstyle{\big |\beta (l)\big >}$}}
\put(34.89,59.78){\circle*{2.22}}
\put(75.11,45.11){\circle*{2.22}}
\put(24.89,70.00){\circle{2.22}}
\put(-190.92,16.93){\circle*{0.00}}
\special{em:linewidth 0.05pt}
\emline{10.00}{10.00}{57}{24.89}{20.00}{58}
\emline{9.78}{34.67}{59}{25.33}{45.11}{60}
\emline{25.33}{45.11}{61}{50.00}{45.11}{62}
\emline{25.11}{45.11}{63}{25.11}{68.89}{64}
\emline{25.11}{45.33}{65}{25.11}{20.22}{66}
\emline{49.78}{45.33}{67}{64.89}{54.89}{68}
\emline{64.89}{54.89}{69}{89.78}{54.89}{70}
\emline{64.89}{80.00}{71}{64.89}{54.89}{72}
\emline{90.22}{30.00}{73}{75.11}{30.00}{74}
\emline{64.89}{54.89}{75}{64.89}{45.33}{76}
\put(64.89,54.67){\circle{2.22}}
\put(62.67,56.44){\makebox(0,0)[rc]{$
\scriptstyle{\big |\alpha (l)\big >}$}}
\emline{25.11}{19.78}{77}{34.89}{19.78}{78}
\emline{49.78}{45.11}{79}{52.22}{42.67}{80}
\emline{53.33}{41.11}{81}{56.00}{38.44}{82}
\emline{57.11}{37.11}{83}{59.56}{34.44}{84}
\put(54.78,39.66){\vector(1,-1){0.2}}
\emline{54.52}{39.91}{85}{54.78}{39.66}{86}
\end{picture}
\end{center}

\noindent
The constant time slice $m=0$ is the chain of vertices
$(\ldots \,, A_{l-1}, A_{l}, A_{l+1},\ldots )$.
The dashed line shows the translation
generated by the $M$-operator (\ref{q13}).

\noindent
{\bf Remark}\,\,The $L$-operator
${\cal L}_l (z)$ (\ref{q12}) has two
degeneracy points: $z=k$ and $z=kq^{-1}$. At the first one the
$R$-matrix is proportional to the permutation operator
${\cal P}$ in ${\bf C}^2\otimes {\bf C}^2$:
$R(1;q,\xi ')=(q-q^{-1}){\cal P}$. Therefore,
\beq
{\cal L}_l (k)=\l _{pn}^{\frac{1}{2}}\Lambda
\,\frac{ \big |\a (l)\big >\,\big <\b (l)\big |}
{\t _{-l}^{p+1}(l,l)\t _{-l+1}^{p}(l,l)}
\label{q15}
\eeq
(cf. (\ref{r4})). At the second one we get:
\beq
{\cal L}_l (kq^{-1})=\l _{pn}^{\frac{1}{2}}\Lambda
\,\frac{ \big |\tilde \b (l)\big >\,\big <\tilde \a (l)\big |}
{\t _{-l}^{p+1}(l,l)\t _{-l+1}^{p}(l,l)}\,,
\label{q16}
\eeq
$$
\big |\tilde \b (l)\big >=
\left ( \begin{array}{cc}
0&-\xi ' \\
1&0 \end{array} \right )
\big |\b (l)\big >\,,
\;\;\;\;\;\;\;\;
\big |\tilde \a (l)\big >=
\left ( \begin{array}{cc}
0&1 \\
-(\xi ')^{-1} &0 \end{array} \right )
\big |\a (l)\big >\,.
$$
Formulas of the type (\ref{q12}), (\ref{q13}) for the
same $L$-$M$-pair can also be written in terms of the vectors
$\big |\tilde \a (l)\big >$,
$\big |\tilde \b (l)\big >$.

\section{Continuous time limit}

In this section we show that the $r$-matrix formula for
local $M$-operators (\ref{r6}) is a degenerate case of
our $R$-matrix formula. It follows from Sect.\,2 that
$\l _{uv}$ plays the role of
lattice spacing for the discrete time variable $m$.
At the first glance, the continuous time limit would then imply
$\l _{v} \to \l _{u}$, i.e. $q\rightarrow 1$,
$\xi \to 1$ in the $R(z;q;\xi )$,
so, in agreement with eq.\,(\ref{r6}), we do get
a classical $r$-matrix.
However, this would imply
$\lim _{q\rightarrow 1}\big |\check \b _{l}\big >=
\big |\a _{l}\big >$ that is certainly wrong in general.
The correct limit to a continuous time coordinate
is more subtle and needs
some clarification.

\subsection{Limiting form of the $M$-operator}

We need a continuous time limit such that the $L$-operator
(\ref{q12}) remains the same.
The naive limit does not work because varying
$\l _{v}$ one changes the STL
and, therefore, the $L$-operator itself.
In the correct limiting procedure, the time lattice spacing
should approach zero independently of $\l _{u}$ and $\l _{v}$.

To achieve the goal,
we exploit the advantage of Miwa's approach \cite{Miwa}
and introduce $w$ --
another "copy" of the chiral flow
$v$ with Miwa's variable $\l _{w}$. The
bilinear equations from Sect.\,2
apparently hold true for $w$ and $\l _{w}$
in place of $v$ and $\l _{v}$ respectively.
Now we can safely tend $\l _{w}\to \l _{u}$
thus eventually getting a continuous flow,
leaving $\l _{v}$ and all the other parameters unchanged. Set

\beq
\hat q=\left (\frac{\l _{wn}\l _{pu}\l _{p'u}}
{\l _{un}\l _{pw}\l _{p'w}}
\right )^{\frac{1}{2}},
\;\;\;\;\;\;\;
\hat \xi =\left (\frac{\l _{un}\l _{pu}\l _{p'w}}
{\l _{wn}\l _{pw}\l _{p'u}}
\right )^{\frac{1}{2}},
\;\;\;\;\;\;\;\;
-\l _{uw}\equiv \varepsilon
\label{lim1}
\eeq
and consider the limit
$\varepsilon \to 0$. We have:
\beq
\hat q =1+
\frac{\l _{p'u}\l _{un}\!+\!\l _{pu}\l _{p'n}}
{2\l _{un}\l _{pu}\l _{p'u}}\,
\varepsilon +O(\varepsilon ^{2})\,,
\;\;\;\;\;\;\;\;
\hat \xi =1+
\frac{\l _{p'u}\l _{un}\!-\!\l _{pu}\l _{p'n}}
{2\l _{un}\l _{pu}+\l _{p'u}}\,
\varepsilon +O(\varepsilon ^{2})\,.
\label{lim2}
\eeq

Discrete $M$-operators are defined up to multiplication
by a scalar function of $z$ independent of dynamical variables.
To pursue the continuous time limit, it is convenient to
normalize the $M$-operators by the condition
${\cal M}_{l}^{0}(z)=I$ at $\varepsilon =0$. Then the next term
(of order $\varepsilon$)
yields local $M$-operator of a continuous time flow.

To implement this project, consider the 4D lattice with
coordinates $(u,v,w,n)$. The $m=0$ slice of the former STL is
embedded into this lattice as the 1D sublattice with vertices
$A_l =(l,l,0,-l)$, $l\in {\bf Z}$.
The continuous time $M$-operator
at the site $A_l$
is obtained by expansion of the discrete $M$-operator
${\cal M}_{\bar B_{l}' \leftarrow A_{l}}(z)
=\l _{un}k(z-z^{-1})I+O(\varepsilon )$ which generates
the translation $A_l =(l, l, 0, -l)$ $\longrightarrow $
$\bar B' =(l+1,l,-1, -l)$. Passing to the normalized
$M$-operator, we have:
\beq
{\cal M}^{0}_{\bar B_{l}' \leftarrow A_{l}}(z)=I +
\varepsilon \tilde M_{l}(z) +O(\varepsilon ^{2})\,.
\label{lim4}
\eeq
For clarity, the $u,w$-section of the
4D lattice ($v=\mbox{const}$,
$n=\mbox{const}$) is displayed in the figure:

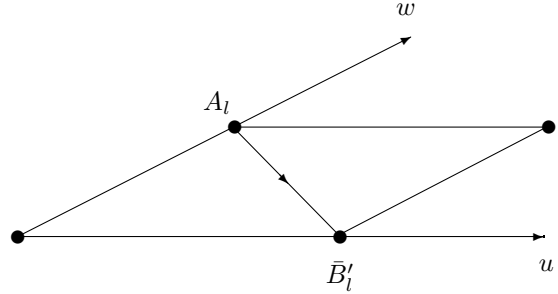
\begin{center}
\special{em:linewidth 0.4pt}
\unitlength 0.7mm
\linethickness{0.4pt}
\begin{picture}(127.50,66.25)(10.00,00.00)
\put(119.33,20.00){\vector(1,0){0.2}}
\emline{19.67}{20.00}{1}{119.33}{20.00}{2}
\emline{60.42}{40.83}{3}{120.00}{40.83}{4}
\emline{120.00}{40.83}{5}{80.00}{20.00}{6}
\emline{60.42}{40.83}{7}{80.83}{20.00}{8}
\put(70.86,30.16){\vector(1,-1){0.2}}
\emline{69.32}{31.77}{9}{70.86}{30.16}{10}
\put(19.58,20.00){\circle*{2.50}}
\put(60.83,40.83){\circle*{2.50}}
\put(120.42,40.83){\circle*{2.50}}
\put(80.83,20.00){\circle*{2.50}}
\put(15.00,12.50){\makebox(0,0)[cc]{$\phantom{a}$}}
\put(80.83,12.50){\makebox(0,0)[cc]{$\bar B_{l}'$}}
\put(57.50,45.42){\makebox(0,0)[cc]{$A_{l}$}}
\put(127.50,45.42){\makebox(0,0)[cc]{$\phantom{a}$}}
\put(120.00,14.50){\makebox(0,0)[cc]{$u$}}
\put(94.17,57.92){\vector(2,1){0.2}}
\emline{19.58}{20.00}{11}{94.17}{57.92}{12}
\put(93.33,63.25){\makebox(0,0)[cc]{$w$}}
\end{picture}
\end{center}

\noindent
The point $\bar B_{l}'$ tends to the point
$A_l$ as $\varepsilon \to 0$, so the parallelogram collapses
to the $u$-axis.
Using eqs.\,(\ref{c3}), (\ref{p'uv}), we obtain:
\beq
\tilde M_{l}(kz)\!=\!\frac{1}{2\l _{p'u}(z\!-\!z^{-1})}
\!\left ( \!\!
\begin{array}{ccc}
\!\displaystyle{\frac{1}{z}}-z+(z+
\displaystyle{\frac{1}{z}})U_l &&
2\mu ^{-2}\!\displaystyle{\frac{\t _{-l}^{p}(l\!+\!1,l)
\t _{\!-l+1}^{p} (l\!-\!1,l)}
{\t _{-l}^{p+1}(l,l)\t _{-l+1}^{p}(l,l)}}\!\\&&\\
2\mu ^{-2}\!\displaystyle{\frac{\t _{-l}^{p+1}(l\!+\!1,l)
\t _{\!-l+1}^{p+1} (l\!-\!1,l)}
{\t _{-l}^{p+1}(l,l)\t _{-l+1}^{p}(l,l)}}&&
\!\!\!2\displaystyle{\frac{z\l _{p'u}}{\l _{un}}}
\!+\!
2\displaystyle{\frac{\l _{p'u}}{z\l _{pu}}}
\!+\!z\!+\!\displaystyle{\frac{1}{z}}
\!-\!(z\!+\!\displaystyle{\frac{1}{z}}\!)U_l\!
\end{array}\!\!\!\right ),
\label{lim5}
\eeq
where
$$
U_l \equiv \frac{\t _{-l}^{p+1}(l+1,l)
\t _{-l+1}^{p}(l-1,l)}
{\t _{-l+1}^{p}(l,l)
\t _{-l}^{p+1}(l,l)}\,.
$$
Note the redundant freedom which allows one to add to the
$\tilde M_l(z)$ a term $h(z)I$ with a scalar function $h(z)$
independent of dynamical variables. This obviously does not
affect the zero curvature condition (\ref{czc}).
In the next subsection
we shall use this freedom to redefine the $M$-operator.

\subsection{Comparison with the $r$-matrix formula}

Rather than to compute the limit of the r.h.s. of eq.\,(\ref{q13})
directly, it is easier to compare eq.\,(\ref{r6})
(for some trial $r$-matrix) with
eq.\,(\ref{lim5}) and look at the "discrepancy" (if any).
As a "first approximation"
let us try the standard classical $r$-matrix

\beq
r^{(0)}(z)= \frac{1}{2(z-z^{-1})}\left [
(z+z^{-1})I\otimes I +
2\sigma _{1}\otimes \sigma _{1}
+2\sigma _{2}\otimes \sigma _{2}
+(z+z^{-1})\sigma _{3}\otimes \sigma _{3}\right ]
\label{r}
\eeq
$$
=\frac{1}{z-z^{-1}}\left (
\begin{array}{ccccccc}
z+z^{-1}&&0&&0&&0 \\ &&&&&& \\
0&&0&&2&&0 \\ &&&&&& \\
0&&2&&0&&0 \\ &&&&&& \\
0&&0&&0&&z+z^{-1} \end{array}
\right )
$$

\noindent
which works in the lattice SG model.
It is skew-symmetric, i.e.
for $z_1 \neq z_2$ it enjoys the property
$r_{12}^{(0)}(z_1 /z_2 )
=-r_{21}^{(0)}(z_2 /z_1 )$
where we use the notation from Sect.\,5.1.

Plugging the vectors (\ref{q11}) into (\ref{r6})
with this $r$-matrix and using eq.\,(\ref{u4}), we compute
the "discrepancy":

\beq
\frac{\l _{pn}\l _{p'n}\!-\!\l _{un}^{2}}
{2\l _{un}\l _{pu}\l _{p'u}}\,
\frac{\big < \b (l)\big |r^{(0)}(z)\big |\a (l-1)\big >}
{\big < \b (l)\big |\a (l-1)\big >}-
\tilde M_{l}(kz) =
\frac{1}{2}\!\left (
\begin{array}{ccc}
\!\l _{p'u}^{-1}\!
&&0 \!\\ && \\
\!0 && \! \l _{pu}^{-1}-\l _{un}^{-1}\!
\end{array}\! \right )
\label{lim6}
\eeq

\noindent
which stands in the right hand side. The result means
that our "first approximation" is not so bad.
To get zero in the r.h.s.,
we need two corrections -- one in the
$\tilde M_{l}(z)$ and one in the $r^{(0)}$.
The first one is rather a matter of definition.
Using the redundant
freedom mentioned above,
we are free to redefine the $M$-operator:
\beq
M_l (z)\equiv \tilde M_{l}(z)+\frac{1}{4}
\left ( \l _{p'u}^{-1}
+\l _{pu}^{-1}-\l _{un}^{-1}\right )I\,.
\label{lim7}
\eeq
The second one is more important. Let
$r^{(\kappa )}(z)$ be the modified $r$-matrix:
\beq
r^{(\kappa )}(z)=r^{(0)}(z)+\kappa I\otimes \sigma _{3}\,,
\label{lim9}
\eeq
\beq
\kappa =
\frac{\l _{p'u}\l _{un}-\l _{pu}\l _{p'n}}
{2(\l _{pn}\l _{p'n}-\l _{un}^{2})}\,.
\label{lim10}
\eeq
The $\kappa$-term violates the skew-symmetry; nevertheless,
the $r^{(\kappa )}(z)$ obeys the
classical Yang-Baxter equation
written in the form modified for not necessarily
skew-symmetric $r$-matrices \cite{BabVia}:
\beq
\left [ r^{(\kappa )}_{12}(z_1 /z_2 ),
r^{(\kappa )}_{13}(z_1 /z_3 ) \right ]\!+\!
\left [ r^{(\kappa )}_{12}(z_1 /z_2 ),
r^{(\kappa )}_{23}(z_2 /z_3 ) \right ]\!+\!
\left [ r^{(\kappa )}_{32}(z_3 /z_2 ),
r^{(\kappa )}_{13}(z_1 /z_3 ) \right ]\!=\!0\,.
\label{cyb}
\eeq

Incorporating the corrections into eq.\,(\ref{lim6}),
we arrive at the $r$-matrix formula

\beq
M_l (kz)=
\frac{\l _{pn}\l _{p'n}\!-\!\l _{un}^{2}}
{2\l _{un}\l _{pu}\l _{p'u}}\,
\frac{\big < \b (l)\big |r^{(\kappa )}(z)\big |\a (l-1)\big >}
{\big < \b (l)\big |\a (l-1)\big >}\,.
\label{lim8}
\eeq

\noindent
At $\kappa =0$ (i.e. $\l _{p'u}\l _{un}=\l _{pu}\l _{p'n}$)
one gets the standard $r$-matrix. (This just happens for the
lattice SG model.) Note that in this case
$\hat \xi =1+O(\varepsilon ^2)$ (see (\ref{lim2})), so the right
classical $r$-matrix is obtained as a "germ" of the quantum one:
$$
r^{(0)}(z)=\lim _{\varepsilon \to 0}
\frac{R(z;\hat q, \hat \xi )-
(z-z^{-1})I\otimes I}{(\hat q -1)(z-z^{-1})}
$$
(provided $\hat \xi =1+O(\varepsilon ^2)$!).

\noindent
{\bf Remark}\,\,The non-skew-symmetric $r$-matrix (\ref{lim9})
signifies that
the $L$-operator at $\kappa \neq 0$ is not ultralocal.
In this case Poisson brackets between
elements of the monodromy matrix ${\cal T}$
are given by a general quadratic
Poisson bracket algebra
$$
\{{\cal T}_1 ,\, {\cal T}_2 \}=
r^{+}_{12}{\cal T}_{1}{\cal T}_{2}
+{\cal T}_{1}s^{+}_{12}{\cal T}_{2}
-{\cal T}_{2}s^{-}_{12}{\cal T}_{1}
-{\cal T}_{1}{\cal T}_{2}r^{-}_{12}
$$
studied (together with its quantum version)
in \cite{FM},\,\cite{NCP},\,\cite{SStSh}\,.
The matrices
$r^{\pm}$, $s^{\pm}$ meet a number of consistency
conditions which ensure antisymmetry and Jacobi identity
for the Poisson bracket. Besides, in integrable systems
the constraint $r^{+}+s^{+}=s^{-}+r^{-}$ holds true.
Under these conditions, the (non-skew-symmetric) $r$-matrix
$r^{+}-s^{-}$ satisfies the Yang-Baxter equation
of the form (\ref{cyb}). In our case the quadruplet
$r^{+}, s^{+}, s^{-}, r^{-}$ is as follows:
$r^{+}=r^{(0)}$,
$s^{+}=-\kappa \sigma _{3}\otimes I$,
$s^{-}=-\kappa I\otimes \sigma _{3}$,
$r^{-}=r^{(0)}+\kappa (I\otimes \sigma _{3}-
\sigma _{3}\otimes I)$, so
$r^{+}-s^{-}=r^{(\kappa )}$.

\section{Conclusion}

The main result of this work is the $R$-matrix representation
(\ref{q12}), (\ref{q13}) of the local $L$-$M$ pair for the
classical discrete HM model.
In our opinion, the very fact that typical quantum $R$-matrices
naturally arise in a purely classical problem is
important and interesting by itself.
To recover them inside classical discrete problems,
one has to pass
to Hirota's bilinear formalism and make use of Miwa's
interpretation of discrete flows.
As a by-product, we
have shown that components of the vectors $\big |\a \big >$,
$\big |\b \big >$ (representing the $L$-operator at the
degeneracy point) are $\t$-functions.

Let us recall that the quantum Yang-Baxter equation already
appeared in connection with purely classical problems,
though in a different
context \cite{Sklyanin85},\,\cite{WX}. However, the class
of solutions relevant to classical
problems is most likely very far from
$R$-matrices of known quantum integrable models.
In our construction, the role of the quantum
Yang-Baxter equation remains obscure; instead,
the most popular 4$\times$4 trigonometric quantum $R$-matrix is
shown to take part in the zero curvature representation of a
classical discrete model. We believe that a conceptual
explanation of this phenomenon nevertheless relies on the
quantum Yang-Baxter equation.

We should stress that the "quantum deformation parameter" $q$
of the trigonometric $R$-matrix in our context
seems to have nothing to do with any kind of quantization.
This fact suggests to extend this hidden $R$-matrix
structure to the quantum level. In this case the vectors
$\big |\a \big >$, $\big |\b \big >$ will have
operator components. If such an
extension does exist, this would mean that there are
{\it two} different $R$-matrices in a quantum integrable
model rather than one. At the same time this would
be an intelligent explanation of our $R$-matrix
formulas -- in the classical
limit one of the $R$-matrices degenerates
to a classical one while the
other one survives.

As far as other possible projects are concerned, there are
many problems apparently arising from this work.
To mention only few, we
point out first of all a detailed comparison with the
hamiltonian approach, which is missing in our exposition.
More specifically, it would be very instructive to identify the
Poisson bracket algebra for elements of the $L$-operator
(\ref{k3}) (in general non-ultralocal)
and, starting from this algebra, to give an
alternative derivation of the $r$-matrix formula
(\ref{lim8}) for the $M$-operator.
Another interesting question is to
find a discrete time analogue of the non-local generating
function (\ref{r3}) of $M$-operators.
At last, we point out the challenging problem to pursue
all the program for the "master model" with 2$\times$2
$L$-$M$-pair -- a discrete analogue of the Landau-Lifshitz
model ($XYZ$ magnet). In this case
the spectral parameter lives on an elliptic curve
and the proper extension of the bilinear formalism
is not completely clear. (In particular, how the
reduction condition (\ref{s1}) looks like?)

\section*{Acknowledgements}

I am grateful to O.Lipan, I.Kri\-che\-ver, A.Vol\-kov and especially
to S.Khar\-chev and P.Wieg\-mann for illuminating discussions
and critical remarks. I also thank T.Asta\-khova for help
in preparing figures. This work was supported
in part by RFBR grant 95-01-01106 and grant 96-15-96455 for
support of scientific schools.

\end{document}